\documentclass{svjour3}

\usepackage{amsmath,amssymb}
\usepackage{proof}
\usepackage{xspace}
\usepackage{url}
\usepackage{graphics} % use for the Pitts-Gabbay quantifier
\usepackage{natbib} % to help with bib
\usepackage{fullpage}

\long\def\ignore#1{}

\newcommand{\lra}{\longrightarrow}
\newcommand{\ra}{\rightarrow}
\newcommand{\lvl}{{\rm lvl}}
\newcommand{\supp}{{\rm supp}}

\newcommand{\hsl}[1]{\hbox{\sl #1}}
\newcommand{\lam}[2]{\hbox{\sl lam} \; #1 \; #2}

\newcommand{\app}[2]{\hbox{\sl app} \; #1 \; #2}
\newcommand{\arr}[2]{\hbox{\sl arr} \; #1 \; #2}
\newcommand{\of}[2]{\hbox{\sl of\/} \; #1 \; #2}
\newcommand{\IH}{\hbox{\sl IH}}
\newcommand{\member}[2]{\hbox{\sl member} \; #1 \; #2}
\newcommand{\nil}{\hbox{\sl nil}}
\newcommand{\seq}[3]{\hbox{\sl seq}_{#1} \; #2 \; #3}

\newcommand{\prog}[2]{\hbox{\sl prog} \; #1 \; #2}
\newcommand{\cntx}[1]{\hbox{\sl cntx} \; #1}

\newcommand{\nat}[1]{\hbox{\sl nat} \; #1}
\newcommand{\name}[1]{\hbox{\sl name} \; #1}
\newcommand{\fresh}[2]{\hbox{\sl fresh} \; #1 \; #2}

\newcommand{\tup}[1]{\langle #1\rangle}
\newcommand{\new}{\reflectbox{\ensuremath{\mathsf{N}}}\xspace}

\newcommand{\tlam}[3]{\lambda #1 \! : \! #2 . #3}
\newcommand{\eval}[2]{\hbox{\sl eval} \; #1 \; #2}
\newcommand{\step}[2]{\hbox{\sl step} \; #1 \; #2}
\newcommand{\sn}[1]{\hbox{\sl sn} \; #1}
\newcommand{\reduce}[2]{\hbox{\sl reduce} \; #1 \; #2}
\newcommand{\zerop}{\hbox{\sl zero}}
\newcommand{\truep}{\hbox{\sl true}}
\newcommand{\falsep}{\hbox{\sl false}}
\newcommand{\succp}[1]{\hbox{\sl succ} \; #1}
\newcommand{\predp}[1]{\hbox{\sl pred} \; #1}
\newcommand{\iszerop}[1]{\hbox{\sl iszero} \; #1}
\newcommand{\ifp}[3]{\hbox{\sl if} \; #1 \; #2 \; #3}
\newcommand{\recp}[2]{\hbox{\sl rec} \; #1 \; #2}
\newcommand{\nump}{\hbox{\sl num}}
\newcommand{\boolp}{\hbox{\sl bool}}

\newcommand{\FOL}{FO\lambda}
\newcommand{\N}{{\rm I} \! {\rm N}}
\newcommand{\FOLDN }{\ensuremath{\FOL^{\Delta\N}}\xspace}
\newcommand{\foldnb}{$FO\lambda^{\Delta\nabla}$\xspace}
\newcommand{\G}{\ensuremath{\cal G}\xspace}
\newcommand{\logic}{\G}
\newcommand{\LG}{$LG^\omega$\xspace}
\newcommand{\ie}{{\em i.e.}}
\newcommand{\eg}{{\em e.g.}}

\newcommand{\hh}{$hH^2$\xspace}

\newcommand{\defL}{\hbox{\sl def}\mathcal{L}}
\newcommand{\defR}{\hbox{\sl def}\mathcal{R}}

\newcommand{\unrhdL}{\unrhd\mathcal{L}}
\newcommand{\unrhdR}{\unrhd\mathcal{R}}
\newcommand{\cut}{\hbox{\sl cut}}
\newcommand{\botL}{\bot\mathcal{L}}

\newcommand{\lorL}{\lor\mathcal{L}}

\newcommand{\landL}{\land\mathcal{L}}

\newcommand{\forallR}{\forall\mathcal{R}}
\newcommand{\nablaL}{\nabla\mathcal{L}}

\newcommand{\existsL}{\exists\mathcal{L}}

\newcommand{\IL}{\mathcal{IL}}
\newcommand{\CIR}{\mathcal{CIR}}

\newcommand{\mueq}{~\stackrel{\mu}{=}~}
\newcommand{\nueq}{~\stackrel{\nu}{=}~}

\newcommand{\cas}[1]{[\![ #1 ]\!]}

\newcommand{\TRUE}{\mbox{TRUE}}
\newcommand{\AND}{\mbox{AND}}
\newcommand{\AUGMENT}{\mbox{AUGMENT}}
\newcommand{\GENERIC}{\mbox{GENERIC}}
\newcommand{\BACKCHAIN}{\mbox{BACKCHAIN}}

\newcommand{\term}[1]{\hbox{\sl term} \; #1}
\newcommand{\pathp}[2]{\hbox{\sl path} \; #1 \; #2}
\newcommand{\uabs}[1]{\hbox{\sl abs} \; #1}
\newcommand{\pdone}{\hbox{\sl done}}
\newcommand{\pleft}[1]{\hbox{\sl left} \; #1}
\newcommand{\pright}[1]{\hbox{\sl right} \; #1}
\newcommand{\bnd}[1]{\hbox{\sl bnd} \; #1}
\newcommand{\ctxs}[2]{\hbox{\sl ctxs} \; #1 \; #2}

\newcommand{\ctx}[1]{\hbox{\sl ctx} \; #1}

\begin{document}
\title{
A two-level logic approach to reasoning about computations
% \\
%  The idea using the term ``computational logic'' is that this
%  suggests both a logic and its supporting implementation: think to
%  Boyer&Moore's book
% A computational logic for reasoning about logic-based specifications\\
% A two level-logic approach to reasoning about computation\\
% Specifying and reasoning using a two level-logic approach
}
\author{Andrew Gacek \and Dale Miller \and Gopalan Nadathur}
\institute{A. Gacek and D. Miller \at
           INRIA Saclay - \^Ile-de-France \& LIX/\'Ecole Polytechnique, Palaiseau, France\\
           \email{gacek at lix.polytechnique.fr, dale.miller at inria.fr}
           \and
           G. Nadathur \at Department of Computer Science and
           Engineering, University of Minnesota\\
           4-192 EE/CS Building, 200 Union Street SE, Minneapolis, MN
	   55455 USA\\
           \email{gopalan at cs.umn.edu}
}
\maketitle

\begin{abstract}
Relational descriptions have been used in formalizing
diverse computational notions, including, for example, operational
semantics, typing, and acceptance by non-deterministic machines.
We therefore propose a (restricted) logical theory
over relations as a language for specifying such notions. Our {\em
  specification logic} is further characterized by an ability to
explicitly treat binding in object languages. Once
such a logic is fixed, a natural next question  
is how we might prove theorems about specifications
written in it.  We propose to use a second logic, called a {\em
  reasoning logic}, for this purpose. 
A satisfactory reasoning logic should be able to 
completely encode the specification logic. Associated
with the specification logic are various notions of binding:
for quantifiers within formulas, for eigenvariables within
sequents, and for abstractions within terms.  To provide a natural
treatment of these aspects, the reasoning logic must encode
binding structures as well as their associated notions of scope,
free and bound variables, and capture-avoiding substitution. Further, to
support arguments about provability, the reasoning logic should
possess strong mechanisms for constructing proofs by induction and
co-induction.   
We provide these capabilities here by using a logic called \logic which
represents relations over $\lambda$-terms via definitions of
atomic judgments,
contains inference rules for induction and co-induction, and
includes a special {\em generic} quantifier.
% written as $\nabla$.
% and a related
% generalization of equality over $\lambda$-terms called {\em
%  nominal abstraction}.
We show how provability in the specification logic can be
transparently encoded in \logic. We also describe an interactive
theorem prover called Abella that implements \logic and this two-level
logic approach and we present several examples that demonstrate the
efficacy of Abella in reasoning about computations.

\end{abstract}

%%% Local Variables: 
%%% mode: latex
%%% TeX-master: "root"
%%% End: 

% LocalWords:  eigenvariables Abella

%\tableofcontents % Just to provide an outline of the paper
\section{Introduction}\label{introduction}

% Specification language - specification logic

We are interested in this paper in specifying computations and
then reasoning about them. 
A range of formalisms have been used as a means for realizing the
first of these objectives. 
For example, the execution semantics of programming languages have
been describe via the $\lambda$-calculus
\citep{reynolds72acm,plotkin76}, the $\pi$-calculus
\citep{milner92mscs}, and abstract machines \citep{landin64}.
A specification formalism that has been particularly successful and
widely
applicable is {\em operational semantics} in both its
``small-step'' version \citep{plotkin81} and its ``big-step'' version
\citep{kahn87stacs}.
Of the many mature and flexible choices that can be made, we pick here 
{\em relational specifications} and their direct encoding as theories
in restricted logics.
This choice allows us to transparently 
encode operational semantics as well as a range of
other notions including, most notably, typing.
Another consequence of our choice is that our specification language
will, in fact, be a specification {\em logic}. More specifically, it will turn
out to be a simple, well understood logic
that can be interpreted as a logic programming language in the style of
$\lambda$Prolog \citep{nadathur88iclp}.

% Reasoning language

After one has picked a language for writing specifications, there is
still a choice to be made about a language for reasoning about them. 
The choices of these two languages are often related. 
If one has selected a specification language relying on, say, process
calculus, then a reasoning language that exploits
bisimulations and congruences would be a natural choice: see, for example,
\citep{sangiorgi94ic}.  If one chooses abstract machines for
specifications then inductive definitions are a natural choice for a
reasoning language. 
In this paper, our reasoning language will be a {\em logic}
that contains standard but powerful mechanisms for induction and
co-induction as well as the $\nabla$-quantifier (both in formulas and
in {\em definitions}) \citep{miller05tocl, gacek09corr}.
Our choice of reasoning logic has a number
of appealing aspects, chief among them being that it is powerful
enough to capture within it many other reasoning
techniques such as bisimulation and inductive definitions.

% Argue that we need two logics

The approach we shall describe in this paper is thus characterized
by the use of {\em two} logics, one for specifying
computations and the other for reasoning about these logic
specifications. 
We pick both of logics to be intuitionistic here but
other choices are also sensible: for example, \cite{mcdowell02tocl} used a
linear logic as a specification language in order to provide 
declarative specifications for a programming language with state.
The logical symbols of these two logics will be
separated in our treatment: in fact, 
provability in the specification logic will be an inductively
defined predicate of the reasoning logic.
Although we distinguish the logics in this way, the term structures
that they use will be identical: in particular, the construction
of terms in both logics will use the same application and
abstraction operations. As a result, term equality in the reasoning
logic will immediately reflect term equality in the specification logic.

% Relationship between logics

In many commonly used approaches, it is problematic to treat
specification-language abstraction through reasoning-logic
abstraction. The reasoning logic often involves
function types that contain recursive functions; this is the case, for
example, in Coq and Isabelle/HOL.  If function abstraction 
at the two levels are identified, function types in the specification
language would also have to contain recursively defined functions. 
Since the specification language is intended to treat syntactic
expressions and not general functions, this raises issues about the
adequate representation of syntax: see, for 
example, the discussion about ``exotic terms'' in
\citep{despeyroux95tlca}.  In the setting that we shall soon unfold,
we get around such problems by making function types in both the 
specification and the reasoning logic weak: while term equality will
still be governed by the rules for $\alpha\beta\eta$-conversion, this
will happen within the simply typed $\lambda$-calculus that 
does not include stronger principles such as recursion.  To recover
the lost strength, we will use inductively and co-inductively defined
{\em predicates} for reasoning about computations. However, at the
predicate level, the specification logic and the reasoning logic 
will be strictly separated. In summary, function types in both logics will
be weak and will be used exclusively to represent syntax that may
contain bindings.  Following \cite{miller00cl}, we shall call this 
style of encoding data with bindings the $\lambda${\em-tree} approach to
abstract syntax.

%GN the comparison with provability logic considered in this para
%talks only about encoding formulas. I have therefore changed
%``object-level provability'' to a more relevant ``object logic''
When we develop the two-level logic approach in detail, the formulas
of the specification logic will become terms 
of the reasoning logic.
%GN this point seems have been made already in the previous para and
%also seems to be orthogonal to the following discussion, unless I
%have missed something.
% but the specification logic will not have any
%reference to the formulas of the reasoning logic.  
This approach to
encoding an object logic within a second logic should be
contrasted to the approach of provability logic (see, for example,
\citep{smorynski04hpl}) where natural numbers are used to denote
syntactic objects of an object-logic and primitive recursive functions
are used to parse and manipulate those objects-cum-natural-numbers.
Our encoding is more direct: {\em both} terms {\em and} formulas
of the specification logic are represented by  terms in the reasoning
logic, with simple types being used to separate terms from
formulas. Moreover, the presence of binding in the terms of the
reasoning logic makes it 
possible to represent quantified formulas in the specification logic
in an immediate and natural manner.

% Advantages of two-level logic
%GN the above line is correct, but the first line seems to have
%confused this with the use of the same term structures. I have
%therefore edited it to keep the focus on what I think it needs to
%be. 
%% There are a number of advantages for having this overlap between these
%% the term structures of these two logics.  
%% First, only mild encoding techniques are needed: for example,
%% quantified specification formula use $\lambda$-abstraction to bind the
%% quantified variable and the instantiation of quantifiers is captured
%% directly using $\beta$-conversion.
There are a number of advantages to the two-level logic approach to
reasoning and the particular realization of it that we discuss in this
paper. First, because of the term structures used in the
reasoning logic, only mild encoding techniques are needed to embed the
specification logic in it: for example, specification-level term
equality is directly captured, quantification in the specification
logic is treated by using $\lambda$-abstraction to bind the 
quantified variable, and the instantiation of quantifiers is realized 
through $\beta$-conversion.
Second, since specifications are written in a logic and since such a logic
typically has meta-theoretic properties (such as cut-admissibility) that 
can be formalized in the reasoning logic, 
powerful techniques become available for reasoning about
descriptions presented in the specification logic.
Third, as a series of examples illustrates, this two-level logic
approach can result in natural, readable, and completely formal proofs
of well-known theorems about computational systems. 
Finally, when one moves to implementing theorem
provers based on this architecture, only one notion of binding,
variable, term equality, substitution, and unification needs to be
treated for both logics.

% Map of paper

In the next section we describe the aspects of the reasoning logic
\logic that we shall use in this paper.
Section~\ref{sec:two-level-reasoning} presents the specification
language \hh and shows how cut-free sequent calculus provability for it
can be given an adequate encoding in \logic.
Section~\ref{sec:abella-architecture} describes briefly the structure
of a theorem-prover called Abella that can be used to interactively
construct sequent calculus proofs in \logic. This description is then
exploited in  
Section~\ref{sec:examples} to present examples of the use of the
two-level logic approach.  Section~\ref{sec:related-work} describes
related work and Section~\ref{sec:conclusion} concludes with an
indication of some future directions.

%%% Local Variables: 
%%% mode: latex
%%% TeX-master: "root"
%%% End: 

\section{The Reasoning Logic}
\label{sec:logic}

%GN When motivating features, use the perspective that in this paper
%this logic is meant to encode a spec logic that is weaker in the
%sense at least that it does not have treatments of fixed points or
%(co)induction. 
%GN Repeats stuff in the previous section
%We shall use the logic \logic \citep{gacek09corr} to encode a
%specification logic and to prove properties of descriptions written
%in it.  
The logic \logic \citep{gacek09corr} is an extension of an intuitionistic
and predicative subset of Church's Simple Theory of Types
\citep{church40}. Terms in 
\logic are monomorphically typed and are constructed using abstraction
and application from constants and (bound) variables. The provability
relation concerns terms of the distinguished type $o$ that are also
called formulas. Logic is introduced by including special constants
representing the propositional connectives $\top$, $\bot$, $\land$,
$\lor$, $\supset$ and, for every type $\tau$ that does not contain
$o$, the constants $\forall_\tau$ and $\exists_\tau$ of type $(\tau
\rightarrow o) \rightarrow o$.  The binary propositional connectives
are written 
%as usual 
in infix form and the expression $\forall_\tau
x. B$ ($\exists_\tau x. B$) abbreviates the formula $\forall_\tau
\lambda x.B$ (respectively, $\exists_\tau \lambda x.B$).  Type
subscripts are typically omitted from quantified formulas when their
identities do not aid the discussion. 
%We will use a
%shorthand for iterated abstraction and quantification: 
If ${\cal Q}$ is the abstraction operator or a quantifier, 
we will often use the shorthand ${\cal Q}x_1,\ldots,x_n.P$ for the
expression ${\cal Q}x_1\ldots{\cal Q}x_n.P$.

The usual interpretation of universally quantified formulas equates
them with the set of all their instances. However, in (weak)
logics meant for specifications over $\lambda$-tree syntax, 
an expression such as ``$B(x)$ holds for
all $x$'' is often meant as a statement about the existence of a
uniform argument for every instance rather than a more general
assertion about the truth of 
some property for these instances. The $\nabla$-quantifier 
\citep{miller05tocl} is included in \logic to encode such {\it generic}
judgments. Specifically, the language contains  logical constants
$\nabla_\tau$ of  type $(\tau \rightarrow o) \rightarrow o$ for each
$\tau$, not containing $o$, that is in a designated set of {\em
nominal} types.  As with the other quantifiers,
$\nabla_\tau x. B$ abbreviates $\nabla_\tau \lambda x. B$. 

Any adequate notion of derivation must associate with the
$\nabla$-quantifier
at least the idea of generalizing on a unique name, but in
such a way that $\nabla_\tau x. F$ is equivalent to $\nabla_\tau y.
(F[y/x])$; the notation $F[t/x]$ denotes here and below the result of a 
% scope-respecting % The following seems more accurate and standard
capture-avoiding 
replacement of $x$ by $t$ in $F$. The \foldnb
logic \citep{miller05tocl} realizes such a view within a sequent
calculus presentation of intuitionistic 
provability by attaching a local signature to each formula in a
sequent. In many reasoning situations, it is useful to strengthen the
interpretation of $\nabla$ by associating with it the $\nabla${\em
  -exchange rule} given by the equivalence $\nabla x.\nabla y. F \equiv 
\nabla y.\nabla x. F$ and the $\nabla${\em -strengthening rule} given
by the equivalence $\nabla x. F \equiv F$, provided $x$ is
not free in $F$ \citep{tiu06lfmtp}. The $\nabla$-strengthening rule
brings with it an ontological commitment to an arbitrary number of
distinct objects at the types over which $\nabla$-quantification is
permitted. This is an acceptable commitment in many applications where
$\nabla$-quantification 
is typically used to represent object-level free variables which are
themselves infinite in number.
The addition of these rules renders both the length of a local
signature and the order of names in it unimportant. These signatures
can therefore be made implicit by distinguishing the variables bound
by them as {\it nominal constants}. It is necessary to recognize,
however, that the particular names used for such constants have
significance only within a single formula and that, in this situation,
the main impact is to ensure that each name refers to a distinct
atomic object.  

The treatment of the $\nabla$-quantifier outlined above was introduced
in the \LG system \citep{tiu06lfmtp} and has been adopted in
\logic. Specifically, an infinite collection of nominal 
constants is assumed for each type at which $\nabla$-quantification
is permitted.  The set of all nominal
constants is denoted by $\mathcal{C}$. These constants are distinct
from (eigen)variables and the usual, non-nominal
constants that we denote by $\mathcal{K}$.  
We define the {\it support} of a term (or 
formula) $t$, written $\supp(t)$, as the set of nominal constants 
appearing in it. 
A permutation of nominal constants is a type preserving bijection
$\pi$ from $\mathcal{C}$ to $\mathcal{C}$ such that $\{ x\ |\ \pi(x)
\neq x\}$ is finite. Permutations are extended to
terms (and formulas), written $\pi . t$, as follows:
\begin{align*}
& \pi.a = \pi(a), \mbox{ if } a \in \mathcal{C} &
& \pi.c = c, \mbox{ if } c\notin \mathcal{C} \mbox{ is atomic} \\
& \pi.(\lambda x.M) = \lambda x.(\pi.M) &
& \pi.(M\; N) = (\pi.M)\; (\pi.N)
\end{align*}
Given two formulas $B$ and $B'$, we write $B \approx B'$ to denote the
fact that there is a permutation $\pi$ such that $B$
$\lambda$-converts to $\pi.B'$. It is easy to see that $\approx$ is an
equivalence relation. Following the earlier discussion, \logic is
designed to preserve provability of sequents with respect to 
replacement of formulas under this relation.

\begin{figure*}[t]
\[
\begin{array}{ccc}
\infer[id]{\Sigma : \Gamma, B \lra B'}{B \approx B'}
& \qquad &
\infer[\cut]{\Sigma : \Gamma, \Delta \lra C}
            {\Sigma : \Gamma \lra B & \Sigma : B, \Delta \lra C}
\\
\noalign{\medskip}
\infer[\forall\mathcal{L}]{\Sigma : \Gamma, \forall_\tau x.B \lra C}
      {\Sigma, \mathcal{K}, \mathcal{C} \vdash t : \tau &
       \Sigma : \Gamma, B[t/x] \lra C} &&
\infer[\forall\mathcal{R},h\notin\Sigma]
      {\Sigma : \Gamma \lra \forall x.B}
      {\Sigma, h : \Gamma \lra B[h\ \bar{c}/x]}
\\
\noalign{\medskip}
\infer[\exists\mathcal{L},h\notin\Sigma]
      {\Sigma : \Gamma, \exists x. B \lra C}
      {\Sigma, h : \Gamma, B[h\; \bar{c}/x] \lra C} &&
\infer[\exists\mathcal{R}]{\Sigma : \Gamma \lra \exists_\tau x.B}
      {\Sigma, \mathcal{K}, \mathcal{C} \vdash t:\tau &
       \Sigma : \Gamma \lra B[t/x]}
\\
\noalign{\medskip}
\infer[\nabla\mathcal{L},a\notin \supp(B)]
      {\Sigma : \Gamma, \nabla x. B \lra C}
      {\Sigma : \Gamma, B[a/x] \lra C} &&
\infer[\nabla\mathcal{R},a\notin \supp(B)]
      {\Sigma : \Gamma \lra \nabla x.B}
      {\Sigma : \Gamma \lra B[a/x]}
\end{array}
\]
\caption{The core rules of \logic: the introduction rules for the
  propositional connectives are not displayed.} 
\label{fig:core-rules}
\end{figure*}

Figure \ref{fig:core-rules} presents a subset of the core rules for
\logic; the standard rules for the propositional connectives have been
omitted for brevity.  Sequents in this logic have the form $\Sigma :
\Gamma \lra C$ where $\Gamma$ is a set of formulas, $C$ is a formula
and the signature $\Sigma$
contains all the free variables of $\Gamma$ and $C$.  In the rules,
$\Gamma, F$ denotes $\Gamma \cup \{F\}$.  In the $\nabla\mathcal{L}$
and $\nabla\mathcal{R}$ rules, $a$ denotes a nominal constant of
appropriate type. In the $\exists\mathcal{L}$ and $\forall\mathcal{R}$
rules, $h$ is an appropriately typed variable not occurring in $\Sigma$,
$\bar{c}$ is a listing of the variables in $\supp(B)$, and $h\;
\bar{c}$ represents the application of $h$ to these constants;
raising, a technique introduced in \citep{miller92jsc}, is used here to
encode the dependency of the quantified variable on $\supp(B)$. The
judgment $\Sigma, \mathcal{K}, \mathcal{C} \vdash t : \tau$ that
appears in the $\forall\mathcal{L}$ and $\exists\mathcal{R}$ rules
enforces the requirement that the expression $t$ instantiating the
quantifier in the rule is a well-formed term of type $\tau$
constructed from the variables in $\Sigma$ and the constants in ${\cal
K} \cup {\cal C}$. Finally, we note that the $id$ rule gives
expression to the richer notion of equality between formulas. 

The notion of {\it substitution} plays an important role in defining
the remaining rules of the logic. As usual, we identify a substitution
$\theta$ as a type-preserving mapping from variables to terms such
that the set $\{x\ |\ x\theta \neq x\}$, the {\em domain} of $\theta$,
is finite. We denote the mapping of a variable $x$ in the domain of a
substitution to the term $t$ by $t/x$. The usual application of a
substitution $\theta = \{t_1/x_1,\ldots,t_n/x_n\}$ to a term $t$
requires paying attention to the scope of binders. In the presence of
the $\lambda$-conversion rules, such an application, that we write as
$t[\theta]$, is given precisely by the term $((\lambda
x_1\ldots\lambda x_n.t)\ t_1\ \ldots\ t_n)$. In \logic, we also have
to pay attention to the fact that a substitution that is determined in
the context of one formula may have to be applied to another formula;
in this case, we must be careful not to confuse the scopes of nominal
constants. Specifically, letting $\pi$ be a permutation of nominal
constants such that $\pi.c$ does not appear in the range of $\theta$
for any $c \in \supp(B)$, the {\it nominal capture avoiding
  application} of the substitution $\theta$ to the formula $B$ is
written as $B\cas{\theta}$ and is defined to be $(\pi.B)[\theta]$.
This definition is ambiguous since many permutations can be chosen for
$\pi$ but the ambiguity is harmless since the result under all
acceptable choices will be equivalent under $\approx$, the intended
notion of equality for formulas.
% Associated with this refined form of substitution we also have a
% notion of composition: the {\it nominal capture avoiding composition}
% of two substitutions $\theta$ and $\sigma$ is written as $\theta
% \bullet \sigma$ and is such that, for any formula $B$, $B\cas{\theta
%   \bullet \rho} = B\cas{\theta}\cas{\rho}$.\footnote{This composition
%   can be given explicitly by first applying a permutation to the range
%   of $\theta$ that renames nominal constants away from those appearing
%   in the range of $\rho$ and then forming the usual composition.}

The logic \logic supports the possibility of recursively defining
atomic judgments. This allows specifications to be directly embedded in
the logic. For example, list membership can be defined by the
following two clauses for {\sl member}:
\begin{align*}
\forall x, \ell.~ \member x (x::\ell) \triangleq \top &&
\forall x, y, \ell.~ \member x (y::\ell) \triangleq \member x \ell
\end{align*}
The part of the clause to the left of $\triangleq$ is called the {\em
  head} while the part to the right is called the {\em body}. The
intuitive reading of a single clause is that if the body is true then
the head is true. Moreover, the reading of the complete set of clauses
for a given predicate, such as {\sl member}, is that the predicate
holds for some arguments just in the case that the predicate with these
arguments matches the head of one of the clauses and the corresponding
instance of the body of that clause is true.

As seen in the example above, the head of a clause can use patterns to
characterize the structure of arguments. We also allow
$\nabla$-quantification to be used in the head to constrain the
structure of terms relative to nominal 
constants. For instance, the clause $(\nabla z. \name z) \triangleq
\top$ defines a predicate {\sl name} which holds only on nominal
constants. When a clause has both $\forall$ and $\nabla$
quantification, the order of these quantifiers allows us to further
restrict the structure of terms. For example, the clause $\forall x.
(\nabla z. \fresh z x) \triangleq \top$ defines a predicate {\sl
  fresh} which holds only when its first argument is a nominal
constant which does not occur in its second argument. This idea is
particularly useful in recursive definitions such as the following
definition of {\sl cntx} which recognizes well-formed typing contexts:
\begin{align*}
\cntx \nil \triangleq \top &&
\forall \alpha, \ell. (\nabla x. \cntx (\tup{x, \alpha} :: \ell))
\triangleq \cntx \ell
\end{align*}
These clauses say that $\cntx L$ holds if and only if $L$ is a list of
pairs of the form $\tup{x,\alpha}$ in which $x$ is a nominal constant
that does not appear elsewhere in the list.

Formally, definitions consist of a finite set of clauses of the form
$\forall \bar{x}. (\nabla \bar{z}. p\ \bar{t}) \triangleq B$ where $p\
\bar{t}$ and $B$ are formulas, neither of which contain any nominal
constants. Moreover, the free variables of $\bar{t}$ must be among
$\bar{z}, \bar{x}$ and the free variables of $B$ must be among
$\bar{x}$. The logic \G is parameterized by the set of clauses, called
$\mathcal{D}$, chosen for a particular reasoning task. Given a
definitional clause $\forall \bar{x}. (\nabla \bar{z}. p\ \bar{t})
\triangleq B$ and a substitution $\sigma$ such that
the list $\bar{z}\sigma$ contains only distinct nominal constants
which do not appear in $\supp(\bar{x}\sigma)$
and such that the free variables of $B[\sigma]$
are a subset of the free variables of $(p\ \bar{t})[\sigma]$, we say
that $(p\ \bar{t})[\sigma] \triangleq B[\sigma]$ is an {\em instance}
of the original clause. Note that instances do not need to be ground
and may contain other free variables. To treat definitions in our
calculus, we add the rules $\defL$ and $\defR$ shown in
Figure~\ref{fig:def-rules} for unfolding predicates on the left and
the right of sequents using their defining clauses. The expression
$\Sigma \theta$ in the $\defL$ rule, denoting the application of a
substitution $\theta=\{r_1/x_1,\ldots,r_n/x_n\}$ to the signature
$\Sigma$, is defined to be the result of removing from $\Sigma$ the
variables $\{x_1,\ldots,x_n\}$ and then adding every variable that is
free in any term in $\{r_1,\ldots,r_n\}$. This rule also uses the
nominal capture avoiding application of a substitution to a set of
formulas that is defined in the obvious way:
$\Gamma\cas{\theta}=\{B\cas{\theta}\;|\; B\in\Gamma\}$.

\begin{figure}[t]
\begin{equation*}
\infer
 [\defR,
  p\ \bar{t} \triangleq B
   \mbox{ an instance from $\mathcal{D}$}]
 {\Sigma : \Gamma \lra p\ \bar{t}}
 {\Sigma : \Gamma \lra B}
\end{equation*}
\begin{equation*}
\infer
 [\defL]
 {\Sigma : \Gamma, p\ \bar{t} \lra C}
 {\left\{
  \Sigma\theta : \Gamma\cas{\theta}, B \lra C\cas{\theta}
    \;|\;
  \hbox{for all $\theta$ and all instances
    $(p\ \bar{t})\cas{\theta} \triangleq B$
    from $\mathcal{D}$}\right\}}
\end{equation*}
\caption{Definition rules}
\label{fig:def-rules}
\end{figure}

In the $\defL$ rule we consider all possible substitutions which allow
an atomic formula to match the head of a clause in
$\mathcal{D}$. Note that these substitutions are intended to affect
the eigenvariables $\Sigma$. For example, consider applying the
$\defL$ rule to the sequent \[\Sigma, x, \ell : \Gamma, \member x \ell
\lra C\] assuming the clauses shown earlier for {\sl member}. Two of
the upper sequents for such an application will be the following:
\begin{equation*}
\Sigma, x, \ell' : \Gamma\cas{x::\ell'/\ell}, \top \lra
C\cas{x::\ell'/\ell}
\end{equation*}
\begin{equation*}
\Sigma, x, y, \ell' : \Gamma\cas{y::\ell'/\ell}, \member x \ell'
\lra C\cas{y::\ell'/\ell}
\end{equation*}
The first of these results from the eigenvariable $\ell$ being
replaced by $x::\ell'$ for some new eigenvariable $\ell'$ and the
second corresponds to $\ell$ being replaced by $y::\ell'$ where $y$
and $\ell'$ are new eigenvariables. Note also that these are not the
only upper sequents for the described rule: there will, in fact, be
infinitely many other upper sequents, obtained by choosing more
specific substitutions for the variables in $\Sigma, x, \ell$.

The $\defL$ rule may have no premises. This happens if there
are no substitutions under which an atom in the left of a sequent
matches the head of a clause in $\mathcal{D}$, something that would be
the case if, for example, $\member x nil$ appeared there. In this
case, the rule
provides an immediate proof of its conclusion. At the other
extreme, there may be an infinite number of substitutions which yield
relevant instances as we have just seen. Having an infinite set of
premises is an obstacle to the effective application  
of the rule. However, the following fact about \logic helps overcome
this difficulty in practice: the provability of $\Sigma : \Gamma \lra
C$ implies the provability of $\Sigma\theta : \Gamma\cas{\theta} \lra
C\cas{\theta}$ for any $\theta$. Thus, the set of premises to be
considered can be limited if we can identify a set of most general
upper sequents from which all other upper sequents can be derived by
applying a nominal capture-avoiding substitution. Looking back at the
example of $\Sigma, x, \ell : \Gamma, \member x \ell \lra C$, the two
upper sequents that we have presented explicitly constitute a most
general set of upper sequents for the application of $\defL$ in this
case. In practice, such most general upper sequents are almost 
always computable and finite \citep{gacek09corr}. We do not discuss
these aspects which are important to implementations any further here,
but we will use the general observations to limit consideration in
particular examples to finite sets of most general upper sequents.

Identifying what constitutes a most general upper sequent for the
$\defL$ rule may require some thought in the case of definitions with
$\nabla$-quantification in the head. Consider, for example, the
derivation of the sequent $x, y, 
z : \fresh x y \lra q\ x\ y\ z$ using the $\defL$ rule, assuming that
{\sl fresh} is defined by $\forall y. (\nabla z. \fresh z y)
\triangleq \top$ and $q$ is some predicate. The following two sequents
in which $a$ is a nominal constant would be upper sequents in this
case: 
\begin{align*}
y, z : \top \lra q\ a\ y\ z &&
y, z': \top \lra q\ a\ y\ (z'\ a)
\end{align*}
The second sequent here is strictly more general than the first: we
can obtain the first from the second via the substitution $\{(\lambda
x.z) / z'\}$ while there is no nominal capture-avoiding substitution
which yields the second from the first. In fact, the second sequent
constitutes a complete set of most general upper sequents for the use 
of $\defL$ in this case. Intuitively, in order to obtain such a most
general sequent, the eigenvariables in the lower
sequent must be raised over the nominal constants introduced by the
definition that is used in conjunction with the $\defL$ rule. 
Notice also that the constraints expressed by the quantification in
the head of a clause may necessitate the ``pruning'' of some of such
raising substitutions. For example, while $y$ may be replaced
initially by $y'\ a$ in the sequent $x, y,
z : \fresh x y \lra q\ x\ y\ z$, the need to match the resulting atom
$\fresh x (y' a)$ with the instance $\fresh a y$ of the head of
the clause under the proviso that $y$ cannot depend on $a$ will result
in $y'$ being substituted for by $\lambda u.y$. 
%GN This was rather confusing. I thought it better to explain it by
%saying raising happens uniformly but then may be altered by matching
%with the head of a clause.
%using the rule in obtaining such a of a most general sequent
%must be raised over the nominal constants which are introduced from
%the $\nabla$-quantifiers in the head of a definition except as
%restricted by the definition, \eg, $y$ is not raised over $a$ in the
%sequents above since $\fresh a y$ must hold.

% TODO: Mention 'nominal abstraction' in reference to
%  gacek09corr and add a short discussion about how our presentation
%  is equivalent.

\begin{figure}[t]
\begin{equation*}
\infer[\IL]
      {\Sigma : \Gamma, p\; \bar{t} \lra C}
      {\{\bar{x}_i : B_i[S/p] \lra \nabla \bar{z}_i. S\ \bar{t}_i\} &
       \Sigma : \Gamma, S\; \bar{t} \lra C}
\end{equation*}
\begin{center}
provided $p$ is defined by the clauses
  $\{\forall \bar{x}_i. (\nabla \bar{z}_i. p\ \bar{t}_i) \mueq B_i\}$
and \\
$S$ is a term with no nominal constants and of the same type as $p$.
\end{center}
\caption{The induction rule}
\label{fig:ind-rule}
\end{figure}

The meaning of the set of clauses for a predicate is given by any one
of the possible fixed-points that can be associated with the clauses.
While the $\defL$ and $\defR$ rules do not discriminate between the
fixed points, \logic allows for a refinement that selects the least or
the greatest fixed point, based on an inductive or co-inductive
reading of the clauses for a given predicate. More precisely an {\it
  inductive} clause is denoted by $\mueq$ in place of $\triangleq$
while a {\it co-inductive} clause is denoted by $\nueq$ in place of
$\triangleq$. We require that the clauses for a given predicate be
uniformly annotated to be inductive, co-inductive or neither. The
$\defL$ and $\defR$ rules may be used with clauses in any of these
forms. Predicates that are inductively defined admit additionally the
induction rule $\IL$ shown in Figure~\ref{fig:ind-rule}. In a proof
search setting, the term corresponding to the schema variable $S$ in
this rule functions like the induction hypothesis and is accordingly
called the invariant of the induction. Note that each clause results
in an additional upper sequent for this rule which requires that
clause to preserve the induction hypothesis.
There is also a co-induction rule in the logic \G though it does not
have a natural presentation with the clause-based treatment of
definitions used in this paper \citep{gacek09corr}.

The interpretation of definitions as fixed-points and the possibility
of reading individual clauses inductively or co-inductively is
sensible only if such clauses satisfy suitable stratification
conditions. For example, a clause such as $a\triangleq
(a\supset \bot)$, in which a predicate has a negative dependency on
itself should be forbidden. In this paper, we shall rely on a simple
method for ensuring stratification that is due to
\cite{tiu.momigliano}. This method uses the idea of associating with
each predicate $p$ a natural number, $\lvl(p)$, that is called its
{\em level}.  This measure is then extended to
formulas as follows:
$\lvl(\top) = \lvl(\bot) = 0$;
$\lvl(p\ \bar{t}) = \lvl(p)$;
$\lvl(B \land C) = \lvl(B \lor C) = \max(\lvl(B),\lvl(C)$;
$\lvl({\cal Q} x.B) = \lvl(B)$ where ${\cal Q}$ is $\nabla$,
$\forall$ or $\exists$;
and
$\lvl(B \supset C) = \max(\lvl(B)+1,\lvl(C))$.
In this context, we consider a definition
to be stratified if we can assign levels to predicates in such a way
that for any clause for $p$ with body $B$ in the definition it is the
case that $\lvl(B[\lambda\bar{y}.\top/p]) < \lvl(p)$. The logic \logic
has been shown to be consistent under this constraint
\citep{gacek09corr}.

\section{The Two-level Logic Approach to Reasoning}
\label{sec:two-level-reasoning}

The logic \logic has significant expressive power, being able to treat
$\lambda$-tree syntax directly and to support
inductive and co-inductive reasoning. As such, it can already be used 
for constructing specifications of computations and then for reasoning
about them.  However, we will not use it in this immediate fashion,
choosing instead to embed a specification logic into it and then using
the specification logic to encode the systems that we wish to
formalize. The particular specification logic that we will use in this
scheme is the intuitionistic theory of second-order hereditary
Harrop formulas that we call \hh. This logic provides a
convenient vehicle for formulating structural, rule-based
characterizations of a variety of properties such as evaluation
and type assignment. Informally, one may think of \hh
as an extension of a simple Prolog-like logic with support for
representing and manipulating $\lambda$-tree syntax \citep{miller00cl}.
An especially useful feature of encodings in \hh is that derivations
that are constructed in \hh based on such encodings end up reflecting
the structure of computations in the object systems.\footnote{Since \hh
  is a subset of the $\lambda$Prolog language
  \citep{nadathur88iclp}, these specifications can
  also be compiled and executed, using an implementation
  of $\lambda$Prolog such as Teyjus
  \citep{nadathur99cade,teyjus.website}.} The embedding of \hh within
\logic that we describe transparently reflects derivations in \hh and
hence gives us the ability to formalize a process of
reasoning directly about computations. Moreover, by proving
meta-theoretic properties of \hh within \logic, we obtain a
collection of general logical principles that can be applied in
arguments about computations in any of the encoded object systems.

This section elaborates the specific two-level logic approach outlined
above. 
%GN this part is already clear from the preceding para.
%using \logic
%as the reasoning logic and \hh as the specification
%logic. 
Section~\ref{sec:spec-logic} presents the logic 
\hh and Section~\ref{ssec:two-level-example} describes an example  
specification in \hh.
Finally, Section~\ref{sec:hhencoding} provides an
embedding of \hh into \logic and shows how some of the meta-theory of
\hh can be formalized through this embedding. 

\subsection{The Specification Logic}
\label{sec:spec-logic}

%GN we need to explain how abstraction and application become
%available later for the examples. There are two possible ways to do
%this. I have picked one and left the other in comments.
%As with \logic, the terms of \hh are monomorphically typed and
%constructed using application and abstraction from constants and
%variables. 
Formulas in \hh are of two kinds.  The {\em goal formulas} are
determined by the grammar
\[
G \;=\; \top \mid A \mid G \land G \mid A \supset G \mid
        \forall_\tau x. G,
\]
where $A$ denotes atomic formulas whose arguments are
monomorphically typed $\lambda$-terms and $\tau$ ranges over types that
do not themselves contain the type of formulas.  
% A notable restriction
% on implication in goal formulas is that the left hand side must be an
% atomic formula.  
{\em Definite clauses} are formulas of the form
$\forall x_1 \ldots \forall x_n . (G_1 \supset \cdots \supset G_m
\supset A)$, where $n$ and $m$ may both be zero and where
quantification is, again, over variables whose types do not
contain that of formulas.  This restricted set
of formulas is ``second-order'' in that to the left of an
implication in a definite formula one finds goal formulas and to the
left of an implication in a goal formula, one finds only atomic
formulas.  These definite clauses, in fact, coincide with the
second-order fragment of higher-order hereditary Harrop formulas
\citep{miller91apal}.

% The semantics of \hh is formalized by means of a proof-theoretic
% presentation of what it means for a goal to follow from a set of
% definite clauses. Specifically, we will be concerned with the
% derivation of sequents of the form 

Provability in \hh is formalized by a sequent calculus proof system in
which sequents are of the form  $\Sigma : \Delta \vdash G$, where
$\Delta$ is a list of definite clauses, $G$ is a goal 
formula, and $\Sigma$ is a set of eigenvariables.  The inference rules
for \hh are presented in 
Figure~\ref{fig:hh-rules}: an immediate consequence of the results in
\citep{miller91apal} is that this proof system is complete for the
intuitionistic theory of \hh.
The GENERIC rule introduces an
eigenvariable when read in a proof search direction, and there is
an associated freshness side-condition: $c$ must not already be in
$\Sigma$. In the BACKCHAIN rule, for each term $t_i \in \bar{t}$
we enforce the type constraint that $\Sigma \vdash t_i : \tau_i$
holds where $\tau_i$ is the type of the quantified variable $x_i$.
An important property to note about these rules is that if we use
them to search for a proof of the sequent $\Sigma : \Delta \vdash
G$, then all the intermediate sequents that we will encounter will
have the form $\Sigma' : \Delta, \mathcal{L} \vdash G'$ for some
$\Sigma'$ with $\Sigma \subseteq \Sigma'$, some goal formula $G'$,
and some
list of atomic formulas $\mathcal{L}$. Thus the initial context
$\Delta$ is {\em global}: changes occur only in the list of atoms
on the left and the goal formula on the right. In presenting
sequents, we will elide the signature when it is inessential to the
discussion. 

\begin{figure}[t]
\centering
\begin{equation*}
\infer[\TRUE]
      {\Sigma : \Delta \vdash \top}
      {}
\hspace{1cm}
\infer[\AND]
      {\Sigma : \Delta \vdash G_1 \land G_2}
      {\Sigma : \Delta \vdash G_1 &
       \Sigma : \Delta \vdash G_2}
\end{equation*}
\begin{equation*}
\infer[\AUGMENT]
      {\Sigma : \Delta \vdash A \supset G}
      {\Sigma : \Delta, A \vdash G}
\hspace{1cm}
\infer[\GENERIC]
      {\Sigma : \Delta \vdash \forall_\tau x.G}
      {\Sigma \cup \{c\!:\!\tau\} : \Delta \vdash G[c/x]}
\end{equation*}
\medskip
\begin{equation*}
\infer[\BACKCHAIN]
      {\Sigma : \Delta \vdash A}
      {\Sigma : \Delta \vdash G_1[\bar{t}/\bar{x}] &
       \cdots &
       \Sigma : \Delta \vdash G_n[\bar{t}/\bar{x}]}
\end{equation*}
where $\forall \bar{x} . (G_1 \supset \ldots \supset G_n \supset
A') \in \Delta$ and $A'[\bar{t}/\bar{x}] = A$ 
\caption{Derivation rules for the \hh logic}
\label{fig:hh-rules}
\end{figure}

\begin{figure}
\[
\infer{\Gamma \vdash x : a}{x : a \in \Gamma} \hspace{.5cm}
\infer{\Gamma \vdash m\; n : b}{\Gamma
  \vdash m : (a \to b) & \Gamma \vdash n : a} \hspace{.5cm}
\infer[\mbox{$x$ not in $\Gamma$}]{\Gamma \vdash (\tlam
  x a r) : (a \to b)}{\Gamma, x : a \vdash r : b}
\]
\caption{Rules for relating a $\lambda$-term to a simple type}
\label{fig:typing}
\begin{center}
\begin{tabular}{c}
$\forall m, n, a, b.(\of m (\arr a b) \supset
    \of n a \; \supset \; \of{(\app m n)} b)$\\
$\forall r, a, b.(\forall x.(\of x a  \supset 
    \of{(r \; x)}{b}) \supset \of{(\lam a r)}{(\arr a b)})$
\end{tabular}
\end{center}
\caption{Second-order hereditary Harrop formulas (\hh) encoding simply typing}
\label{fig:hhtyping}
\end{figure}

% DM: Changed this from subsubsection to subsection.  There was a 3.1.1
% but no 3.1.2...  I think it's better to have it as 3.2.

\subsection{An Example}
\label{ssec:two-level-example}

We briefly illustrate the ease with which type assignment for the
simply typed $\lambda$-calculus can be encoded in \hh. There are
two classes of objects in this domain: types and terms. For types
we will consider a single base type called $i$ and the arrow
constructor for forming function types. Terms can be variables
$x$, applications $(m\; n)$ where $m$ and $n$ are terms, and typed
abstractions $(\tlam x a r)$ where $r$ is a term and $a$ is the
type of $x$. The standard rules for assigning types to terms are
given in Figure \ref{fig:typing}. Object-level simple types and
untyped $\lambda$-terms can be encoded in a simply typed
(meta-level) $\lambda$-calculus as follows. We assume the types
$ty$ and $tm$ for representing object-level simple types and
untyped $\lambda$-terms. The simple types are built from the two
constructors $i : ty$ and $\hsl{arr} : ty \to ty \to ty$ and terms
are built using the two constructors $\hsl{app} : tm \to tm \to
tm$ and $\hsl{lam} : ty \to (tm \to tm) \to tm$. Here, the
constructor {\sl lam} takes two arguments: one for the type of the
variable being abstracted and the other for the actual
abstraction. Note, in particular, that the bound variable in an
object-level abstraction will be encoded by an explicit, specification
logic abstraction: thus, the object-level term $(\tlam f {i\to i}
(\tlam x i (f\; x)))$ will be represented by the specification logic
term $\lam {(\arr i i)} (\lambda f. \lam i (\lambda x. \app f
x))$.
% We will be concerned only with closed terms in the object
% language and will therefore not need an explicit (meta-level)
% constructor for variables.
% AG: My issue with the previous sentence is that even if we
% wanted to deal with open terms we would not use an explicit
% constructor. We would just denote them by variables which would
% be bound by some signature \Sigma in a derivation. Rather than
% try to spell this out, I've just avoided the sentence.

Given this encoding of the untyped $\lambda$-calculus and simple
types, the inference rules of Figure~\ref{fig:typing} can be
specified by the \hh definite clauses in Figure \ref{fig:hhtyping}
involving the binary predicate {\sl of}. Note
that this specification in \hh does not maintain an explicit
context for typing assumptions but uses hypothetical judgments
instead. Also, the explicit side-condition in the rule for typing
abstractions is not needed since it is captured by the freshness
side-condition of the GENERIC rule in \hh.

The properties that we prove in \logic will eventually be about
specification logic judgments. To reflect such properties into
related properties about the object system, we will establish
two results about our encodings:
that there exists a bijection, $\phi$, between expressions of
the object system and their specification logic
representations and that this bijection preserves the judgments
of interest. These properties constitute what is referred to as the
{\it adequacy} of an encoding. We illustrate below the structure of
adequacy arguments in the context of our encoding of the simply typed 
$\lambda$-calculus. 

We start by defining the mapping $\phi$ from object-level simple
types to \hh terms of type $tp$ and from object-level
untyped $\lambda$-terms to \hh terms of type $tm$.
\[
\phi(i) = i \qquad
\phi(a \to b) = \arr {\phi(a)} \phi(b)
\]
\[
\phi(x) = x \qquad
\phi(m\ n) = \app {\phi(m)} \phi(n) \qquad
\phi(\tlam x a r) = \lam {\phi(a)} (\lambda x.\phi(r))
\]
In the first case for the mapping of terms, $x$ is used to denote
both an object-level and a corresponding specification logic variable.
Note that under this mapping bound object-level variables will
correspond to variables bound by $\lambda$'s in the specification
logic, and object-level free variables will correspond (eventually) to
eigenvariables in the specification logic. The mapping $\phi$ is
bijective so long as we only allow eigenvariables at type $tm$.
In later arguments, we will need
the fact that bound variables in both the object system and the
specification logic can be renamed so that, for example, rules
with freshness side-conditions can be correctly applied. It is
important that such object-level and specification logic renamings are
carried out in a consistent fashion. A more general form of this
property is that $\phi$ is compositional with respect to
substitution which can be stated as follows:
\begin{equation*}
\phi(r[x := n]) = \phi(r)[\phi(n)/x]
\end{equation*}
Notice that we have used object-level substitution on the left and
specification logic substitution on the right. This equality can be
proved by induction on the structure of $r$.

We now want to define a mapping from object-level derivations of
typing judgments to derivations in \hh of sequents of the form
$\Delta, \mathcal{L} \vdash \of e t$ where $\Delta$ is a list of
the clauses from Figure~\ref{fig:hhtyping} and $\mathcal{L}$ is a
list of atomic formulas of the form $\of {x_1} {a_1}, \ldots, \of
{x_k} {a_k}$ where each $x_i$ is a unique eigenvariable. Towards this
end, we first 
define the following bijection between a list of typing
assumptions $\Gamma$ from the simply typed $\lambda$-calculus and
a list of atomic formulas of the form described for $\mathcal{L}$.
\begin{equation*}
\phi(x_1 : a_1, \ldots, x_k : a_k) =
\of{x_1}{\phi(a_1)}, \ldots, \of{x_k}{\phi(a_k)}
\end{equation*}
Using this, we can define the mapping for the (atomic) typing
derivation for variables as follows:
\begin{equation*}
\phi\left(\raisebox{-1.5ex}{
  \infer[]
        {\Gamma \vdash x_i : a_i}
        {}
  }\right)
=
\raisebox{-1.5ex}{
   \infer[]
         {\Delta, \phi(\Gamma) \vdash \of{x_i}{\phi(a_i)}}
         {}
   }
\end{equation*}
If the object system typing derivation to which $\phi$ is
applied is correct, then it must be
that $x_i : a_i \in \Gamma$. Thus the right-hand side is an
instance of the BACKCHAIN rule on the clause $\of {x_i} \phi(a_i)$
which is in $\phi(\Gamma)$.

Derivations in the object system that have the typing rule for
applications at the end are mapped in the expected way:
\begin{multline*}
\phi\left(\raisebox{-3.5ex}{
  \infer[]
        {\Gamma \vdash m\ n : b}
        {\deduce{\Gamma \vdash m : a \to b}{\vdots\ \ \ } &
         \deduce{\Gamma \vdash n : a}{\vdots}}
  }\right) =
\raisebox{-3.5ex}{
   \infer[]
         {\Delta, \phi(\Gamma) \vdash \of {\phi(m\ n)} \phi(b)}
         {\phi\left(\raisebox{-2ex}{
           \deduce{\Gamma \vdash m : a \to b}{\vdots\ \ \ }
          }\right)
          &
          \phi\left(\raisebox{-2ex}{
            \deduce{\Gamma \vdash n : a}{\vdots}
          }\right)
         }
  }
\\
= \raisebox{-3.5ex}{
   \infer[]
         {\Delta, \phi(\Gamma) \vdash
            \of {(\app {\phi(m)} \phi(n))} \phi(b)}
          {\raisebox{-1.5ex}{
             \deduce{\Delta, \phi(\Gamma) \vdash
               \of {\phi(m)} {(\arr {\phi(a)} \phi(b))}}
               {\phi(\vdots)}
           }
          &
          \raisebox{-1.5ex}{
            \deduce{\Delta, \phi(\Gamma) \vdash
              \of {\phi(n)} {\phi(a)}}
             {\phi(\vdots)}
          }
         }
  }
\end{multline*}
This is clearly a well-formed instance of the BACKCHAIN rule using
the clause for typing applications in $\Delta$.

In mapping derivations in the object system that have the rule for
typing abstractions at the end, we need to be mindful of the
variable naming restriction and how this is realized in the
specification logic. Suppose we want to define the following mapping:
\begin{equation*}
\phi\left(\raisebox{-3.5ex}{
  \infer[]
        {\Gamma \vdash (\tlam x a r) : a \to b}
        {\deduce{\Gamma, x : a \vdash r : b}{\vdots}}
  }\right)
\end{equation*}
Here we assume that $x$ does not appear in $\Gamma$ so that the naming
restriction is satisfied. We map this to the following specification
logic derivation:
\begin{equation*}
\raisebox{-6ex}{
   \infer[\BACKCHAIN]
     {\Delta, \phi(\Gamma) \vdash
        \of {(\lam {\phi(a)} (\lambda x. \phi(r)))}
            (\arr {\phi(a)} \phi(b))}
     {\infer[GENERIC]
        {\Delta, \phi(\Gamma) \vdash \forall x . (\of x \phi(a)
          \supset \of {((\lambda x. \phi(r))\ x)} \phi(b))}
        {\infer[AUGMENT]
          {\Delta, \phi(\Gamma) \vdash \of x \phi(a)
           \supset \of {\phi(r)} \phi(b)}
          {\deduce[]
            {\Delta, \phi(\Gamma), \of x \phi(a) \vdash
               \of {\phi(r)} \phi(b)}
            {\phi(\vdots)}}}}
  }
\end{equation*}
In the GENERIC rule we overload notation to let $x$ be the eigenvariable
we select. Since it does not appear in $\Gamma$ it will not appear in
$\phi(\Gamma)$, and thus the freshness side-condition on the GENERIC
rule is satisfied. In fact, the naming restriction in the object logic
matches up with the freshness side-condition in the specification
logic exactly as needed.

The inverse of the $\phi$ mapping for typing judgments can be defined
in the expected way, and it can be seen from this that $\phi$ is a
bijection. Therefore our encoding of the typing relation is adequate.

\subsection{Encoding Specification Logic Provability in \logic}
\label{sec:hhencoding}
The definitional clauses in Figure~\ref{fig:seq} encode \hh 
provability in \logic; this encoding is based on ideas from 
\citep{mcdowell02tocl}. 
Formulas in \hh are represented in this setting
by terms of type {\sl form} and we reuse the symbols $\land$, $\lor$,
$\supset$, $\top$, and $\forall$ for constants involving this type in
\logic; we assume that the 
context will make clear which reading of these symbols is meant. The
constructor $\langle \cdot \rangle$ is used to inject atomic formulas
in \hh into specially marked expressions of type {\sl form} in
\logic. 
As we have seen earlier, provability in \hh is
about deriving sequents of the form $\Delta,{\cal L}\vdash G$, where
$\Delta$ is a 
fixed list of definite clauses and $\cal L$ is a varying list of
atomic formulas. Our encoding uses the \logic predicate {\sl prog}
to represent the definite clauses in $\Delta$. In particular, the definite
clause $\forall\bar x.[G_1\supset\cdots\supset G_n\supset A]$ is
encoded as the clause $\forall \bar x. \prog A (G_1 \land \cdots \land
G_n) \triangleq \top$ and particular specifications written in \hh
will be reflected into \logic through corresponding collections of {\sl
  prog} clauses. Sequents in \hh are represented in \logic by means of
atomic formulas of the form $\seq N L G$ where $L$ is a list
encoding the atomic formulas in $\cal L$ and where $G$ encodes the goal
formula. The provability of such sequents in \hh, given by the rules in
Figure~\ref{fig:hh-rules}, leads to the clauses that define {\sl seq}
in Figure~\ref{fig:seq}. The argument $N$ that is written as a
subscript in the expression $\seq {N} L G$ encodes (roughly) the height
of the corresponding \hh derivation and is needed in formalizing
proofs by induction on these heights. This argument has type {\sl nt}
that is endowed with two constructors: {\sl z} of type {\sl nt}
and {\sl s} of type {\sl nt}~$\ra~${\sl nt}. 

\begin{figure}[t]
\begin{align*}
&\seq {(s\; N)} L \top \mueq \top
&& \nat z \mueq \top \\
&\seq {(s\; N)} L (B \land C) \mueq \seq N L B \land \seq N L
C
&& \nat (s\ N) \mueq \nat N \\
& \seq {(s\; N)} L (A \supset B) \mueq \seq N {(A :: L)} B \\
& \seq {(s\; N)} L (\forall B) \mueq \nabla x. \seq N L (B\; x)
&& \member B (B::L) \mueq \top \\
& \seq {(s\; N)} L \langle A \rangle \mueq \member A L
&& \member B (C::L) \mueq \member B L \\
& \seq {(s\; N)} L \langle A\rangle \mueq
  \exists b. \prog A b \land \seq N L b \\
\end{align*}
\caption{Second-order hereditary Harrop logic in \logic}
\label{fig:seq}
\end{figure}

A few remarks are appropriate pertaining to the encoding of \hh
provability. First, note that proofs of universally quantified goal
formulas in \hh are generic in nature. Thus, a natural way to encode 
the proof rule for the (specification-logic) universal quantifier
is to use the $\nabla$-quantifier, as is
done in the clause defining {\sl seq} for this case. 
Second, observe that in proving an implication, the atomic
assumption is added, as would be expected, to the list that is the
second argument of {\sl seq}. 
Third, the last clause for {\sl seq} can be seen to implement
backchaining over a given \hh  specification, stored as {\sl prog}
clauses. The matching of atomic judgments to heads of
clauses is handled by the treatment of definitions in the logic
\logic; thus the last rule for {\sl seq} simply performs this matching
and makes a recursive call on the corresponding clause body. Finally,
observe that the way the natural number (subscript) arguments are
used in the {\sl seq} clauses ensures a correct encoding of
the fact that the premise sequents of a rule in \hh must be shorter
than the derivation of the conclusion sequent. 

With this kind of an encoding, we can now formulate and prove in
\logic statements about what is or is not provable in \hh.
Induction over the heights of derivations may be needed in such
arguments and this can be realized via natural number induction on $N$
in $\seq N L P$, realized using induction over the clauses in
Figure~\ref{fig:seq} defining the {\sl nat} predicate. Notice also
that the $\defL$ rule encodes case analysis in the derivation of an
atomic goal, leading eventually to a consideration of the different
ways in which an atomic judgment may have been inferred in the
specification logic.  

\subsubsection{Formalizing Properties of the Specification Logic}
\label{sec:hh-meta-theory}

Since we have encoded the entire derivability relation of \hh, we can
prove general properties about it in \logic and then use these in 
reasoning about particular specifications. For example, the
following formula, which is provable in \logic, states that the
judgment $\seq n \ell g$ is not affected by permuting,
contracting, or weakening the context $\ell$.
\begin{equation*}
\forall n, \ell_1, \ell_2, g. (\seq n {\ell_1} g) \land (\forall
e . \member e \ell_1 \supset \member e \ell_2) \supset (\seq n
{\ell_2} g)
\end{equation*}
This property can be applied to any specification judgment
that uses hypothetical assumptions. Using it with the encoding of
typing judgments for the simply typed $\lambda$-calculus, for example,
we easily obtain that permuting, contracting, or weakening the typing
context of a typing judgment does not invalidate that judgment.

Two additional properties of our specification logic which are useful
and provable in \logic are called the {\em instantiation} and {\em cut}
properties. The instantiation property recovers the notion of
universal quantification from our representation of the specification
logic $\forall$ using $\nabla$. The exact property is
\begin{equation*}
\forall n, \ell, g. \nabla x. (\seq n {(\ell\ x)} {(g\ x)})
\supset \forall t. (\seq n {(\ell\ t)} {(g\ t)}).
\end{equation*}
Stated another way, although $\nabla$-quantification cannot be
replaced by $\forall$-quantification in general, it can be
replaced in this way when dealing with specification judgments.
The cut property allows us to remove hypothetical judgments using
a proof of such judgments. This property is stated as the formula
\begin{equation*}
\forall n, m, \ell, a, g. (\nat n \land \seq n {\ell} \langle
a\rangle) \land (\nat m \land \seq m {(a::\ell)} g) \supset \exists p.
(\nat p \land \seq p \ell g),
\end{equation*}
which can be proved in \logic. To demonstrate the usefulness of the
instantiation and cut properties, we observe that using these together
with our encoding of typing for the simply typed
$\lambda$-calculus leads to an easy proof of the type substitution
%GN There is a bit of an issue here: we are talking about a property
%of the object system based on something we show in G, but this
%requires the adequacy property discussed in the next subsection.
property, \ie, if $\Gamma, x:a \vdash m : b$ and
$\Gamma \vdash n : a$ then $\Gamma \vdash m[x := n]
: b$.

\subsubsection{Adequacy of the Encoding of the Specification
  Logic}

We are eventually interested in lifting the results we prove about
encodings to related results about the original object systems.
In the two-level logic approach, adequacy proofs of this kind
can be factored through an adequacy result for the encoding of the
specification logic; in the present context, this corresponds to the
adequacy of the encoding of \hh in \logic via the definition of {\sl
  seq} and {\sl prog}.  One benefit of the two-level logic approach
is that adequacy of the encoding of the
specification logic needs to be established only once for all
applications, provided this is properly parameterized by the
embedding of specifications themselves via the
{\sl prog} clauses.  Thus, the important statement of
adequacy for the combination of \hh and \logic is the following:

\begin{theorem}\label{thm:seq-adequacy}
Let $\Delta$ be a list of closed definite clauses, $\mathcal{L}$ a
list of atoms, $G$ a goal formula, and $\Sigma$ a set of
eigenvariables containing at least the free variables of $\Delta$,
$\mathcal{L}$, and $G$.  Suppose that all non-logical
specification logic constants and types are represented by
equivalent constants and types in \logic and let $\psi$ denote the
obvious mapping between formulas in \hh and terms in \logic. Suppose
also that 
specification logic $\forall$-quantification (eigenvariables) and
reasoning logic $\nabla$-quantification (nominal constants) are allowed
only at inhabited types. Then $\Sigma : \Delta, \mathcal{L} \vdash
G$ has a derivation in \hh if and only if $\exists n. \nat n \land
\seq n {\psi(\mathcal{L})} {\psi(G)}$ is provable in \logic with
the clauses for {\sl nat}, {\sl member}, and {\sl seq} as stated
before and the clauses for {\sl prog} as given by the prescribed
encoding of $\Delta$.
\end{theorem}

%GN This is a tough para on the whole, I think, with some subtle
%points. I wonder if it may be useful to elaborate a bit on the typing
%issue here, mainly to go it a bit slower for the ``poor reader.''
%DM: I thought that this paragraph was okay, at least for now.
The proof of this theorem is straightforward and its details are
available in \citep{gacek09phd}. The only interesting point is the
relevance of the condition that specification logic
$\forall$-quantification and 
reasoning logic $\nabla$-quantification are allowed only at inhabited
types. This condition is needed because we have chosen to use a 
shallow encoding of the typing judgment of the specification
logic. That is, rather than encoding an explicit typing judgment for
specification logic terms, we have relied on the typing
judgment of \logic to enforce the well-formedness of terms. Due to
the lack of restrictions on the occurrences of nominal constants,
the typing judgment in \logic is more permissive than the
specification logic typing judgment.
However, as the statement of the theorem
indicates, this difference only manifests itself at uninhabited
types. For inhabited types, the instantiation property of \hh can
be used to remove any ``stray'' nominal constants. A deeper
encoding involving an explicit typing judgment would avoid this
condition, but would also impose additional costs in
terms of reasoning both about and through the encoding. 
In our experience, the shallow encoding has turned out to provide a
good balance in practice. 

The theorem above restricts the
definitions of the predicates 
{\sl nat}, {\sl member}, {\sl seq}, and {\sl prog}, but makes no
explicit reference to other predicates. Indeed, the definitions of
other predicates have no affect on the adequacy of the encoding of the
specification logic. Additionally, \logic may make use of additional
constants and types which are unconnected to the constants and types
used to represent the specification logic without affecting the
adequacy of the encoding.

%%% Local Variables: 
%%% mode: latex
%%% TeX-master: "root"
%%% End: 

% LocalWords:  tt prog arr hh cntx logics intuitionistic Harrop encodings ty tm
% LocalWords:  existentials sequents eigenvariables eigenvariable BACKCHAIN tp
% LocalWords:  backchaining bijection bijective nat formedness co cl nadathur
% LocalWords:  iclp Teyjus cade teyjus apal subsubsection renamings mcdowell nt
% LocalWords:  tocl seq derivability GN para gacek phd

\section{The Architecture of Abella}
\label{sec:abella-architecture}

Abella is an interactive theorem prover for the logic \logic which
incorporates the two-level logic approach to reasoning
\citep{gacek08ijcar,abella.website}.  In this section we briefly describe the
architecture of Abella.  In particular, we illustrate how \logic and
the two-level logic approach are presented to the user within this
system and we introduce terminology and notation that are useful in
the example applications that we consider in the next section.

\subsection{Proof Construction, Tactics, and (Co)Induction}

The high-level structure of Abella is similar to that of most other
tactics-based theorem provers. At any time, the state of the prover is 
represented by a collection of subgoals, all of which need to
be solved for the overall proof to succeed. The user applies a
tactic to a subgoal in order to make progress towards a
completed proof. If we think of a completed proof as a derivation
for a sequent in \logic, then the subgoals correspond to
sequents whose derivations will complete the proof being sought. 
A tactic corresponds in this setting to a scheme
for using the rules of \logic to produce new subgoals whose
derivations can, in turn, be used to produce a derivation of the
subgoal under consideration. 

The tactics in Abella are designed to model natural proof steps.
Some tactics serve to collect related proof rules under a single
name. For example, Abella has a ``case analysis'' tactic which
uses a rule such as $\lorL$, $\landL$, $\botL$, $\defL$,
$\existsL$, or $\nablaL$, depending on the structure of the formula to
which it is applied. Other tactics combine the use of many rules in
tandem. 
For example, Abella has an ``apply'' tactic which takes a lemma or
hypothesis of the form $\forall \bar{x}. H_1 \supset \ldots
\supset H_n \supset C$ and hypotheses $H_1', \ldots, H_n'$ and
tries to find terms $\bar{t}$ such that for each $i \in
\{1,\ldots,n\}$ it is the case that
$H_i' \lra H_i[\bar{t}/\bar{x}]$
can be provided a proof using only the {\sl id} rule. If
successful,  the tactic adds a new hypothesis $C[\bar{t}/\bar{x}]$.

Abella has treatments for induction and co-induction which
simplify much of the work involved in formulating invariants and
co-invariants. We will focus on the treatment of induction here: 
further details of the approach to
induction and co-induction in Abella are available in
\citep{gacek09phd}.  Suppose we have the sequent
\begin{equation*}
\Sigma : p\ \bar{t}, H_1, \ldots, H_n \lra C,
\end{equation*}
where $p$ is inductively defined. The induction tactic can be applied
to this sequent by designating $p\ \bar{t}$ as the 
the {\it induction formula}.  The application of the tactic
is based on the additional formula \[\forall \Sigma . (p\ \bar{t})^*
\supset H_1 \supset \ldots \supset H_n \supset C,\] in which
$\forall\Sigma$ denotes a list of universal quantifiers, one for each
variable in $\Sigma$. This formula, which we call the {\it induction
  hypothesis} and denote by $\IH$, has an occurrence in it of the
induction formula that is annotated with $^*$. The formula annotated
in this way in the induction hypothesis can only be matched by
another formula that has the same annotation. The induction tactic now
transforms the original sequent into
\begin{equation*}
\Sigma : \IH, (p\ \bar{t})^@, H_1, \ldots, H_n \lra C.
\end{equation*}
The atomic formula $p\ \bar{t}$ that has the annotation $^@$ here is treated
as if the annotation is not present, with the exception that when it
is unfolded using a $\defL$ rule any new atoms that are introduced
that have $p$ as their head symbol are annotated with $^*$. 
These formulas that are annotated with $^*$ are treated just like the
formula with the $^@$ annotation except that they are also eligible
to be used with the induction hypothesis.
Thus, viewed intuitively, the induction tactic simply generates 
an induction hypothesis that is usable when the induction formula is
unfolded. This tactic can be seen as the special case
of the use of the $\IL$ rule; a detailed justification is presented
elsewhere \citep{gacek09phd}.
% This is not so easy now that we are using pattern-form
%
%  with the invariant
% \begin{equation*}
% S = \lambda\bar{x}. \forall \Sigma. (\bar{x} = \bar{t}) \supset
% H_1 \supset \ldots \supset H_n \supset C.
% \end{equation*}
% The right upper sequent of this application of the $\IL$ rule 
% has a trivial derivation.   The
% left upper sequent, on the other hand, holds the key to the inductive
% argument and a little analysis shows that a derivation for it can be
% filled out by using a proof of the subgoal generated by the induction
% tactic. 

\subsection{Treatment of the Two-level Logic Approach to Reasoning}

Abella incorporates the two-level logic approach to reasoning
using the specification logic \hh and its encoding via {\sl seq}
and {\sl prog}. Moreover, the actual details of the encoding are
hidden from the user. As we have observed already, \hh is a subset of
the $\lambda$Prolog language. Abella allows \hh specifications to be
written in $\lambda$Prolog syntax, thereby permitting one to reason
about computations based on the same descriptions that are used to 
prototype them. Following this approach also
creates the feeling that one is reasoning directly about \hh
derivations that reflect the encoded computations.

Abella uses specialized syntax to simplify the presentation of
specification logic judgments.
In particular, the judgment $\exists n . \nat n \land \seq n L
\langle A \rangle$ is presented as $\{L \vdash A\}$. Moreover, the
list $L$ is decomposed into a presentable format that matches the
way hypotheses are typically written in an \hh judgment. For
example, the judgment $\{H_1 :: H_2 :: L \vdash A\}$ is presented
more suggestively as $\{L, H_2, H_1 \vdash A\}$. If the list ends in
{\sl nil} rather than a variable then we simply write $\{H_2,
H_1 \vdash A\}$.
If
the entire list is {\sl nil} then we elide even the turnstile, writing
the judgment as $\{A\}$. Looking at the clauses in
Figure~\ref{fig:seq}, we see that any {\sl seq} judgment in which the
last argument is a non-atomic goal can be immediately and
deterministically transformed into a collection of such judgments in
which the last argument is an atomic 
goal. Thus the specialized $\{\cdot\vdash\cdot\}$ notation is the only 
representation of the specification logic that needs to be
exposed to the user.
For example, using the clauses from Figure~\ref{fig:hhtyping} in
Abella, case analysis on an assumption $\{\of {(\lam A R)} (\arr A
B)\}$ results directly in the new assumption $\{\of c A \vdash \of
{(R\ c)} B\}$ where $c$ is a nominal constant.

As we have observed in Section~\ref{sec:hhencoding}, \hh is a logic
with notable meta-theoretic properties which can be formalized and
established as theorems of \logic. Combining such
results with the apply tactic leads to an expanded collection of
tactics within Abella which are geared to reasoning about \hh
specifications.
For 
example, given $\{L, A \vdash B\}$ and $\{L \vdash A\}$ the {\em cut}
tactic allows one to 
derive $\{L \vdash B\}$.  Similarly, given a hypothesis $\{L\vdash
A\}$, a nominal constant $v$ in that hypothesis, and a term $t$ of the
same type as $v$, the {\em inst} tactic allows one to derive
$\{L[t/v] \vdash A[t/v]\}$.  Also, a tactic is available for deriving
from $\{L\vdash A\}$  the hypothesis $\{K\vdash A\}$ if the list $L$
denotes a set that is a subset of the set denoted by the list $K$.

% In particular, the property of \hh which
% allows contexts to be treated like essentially like sets is
% incorporated into many of the tactics in Abella so that one rarely
% needs to directly manipulate contexts in this way.

Finally, the treatment of induction described previously is extended
to formulas of the form $\{L \vdash A\}$ by attaching annotations
directly to such formulas. This treatment is justified by
unfolding $\{L \vdash A\}$ to $\exists n. \nat n \land \seq n L
\langle A\rangle$, applying the $\existsL$ and $\landL$ rules, and
using the induction tactic with $\nat n$ as the induction formula.

%%% Local Variables: 
%%% mode: latex
%%% TeX-master: "root"
%%% End: 

% LocalWords:  Abella OCaml Nadathur subgoals IH ack Ackermann sep CH subgoal
% LocalWords:  instantiatable tuples sequents prog gacek ijcar abella phd
% LocalWords:  Prolog

\section{Examples}
\label{sec:examples}

We now illustrate the two-level logic approach to reasoning
through concrete examples. We start with a specification of
evaluation and typing for the simply typed $\lambda$-calculus
for which we prove some basic properties.
We then consider extensions in two different
directions. In one direction, we enrich the collection of terms 
to the language of PCF
\citep{plotkin77} and we demonstrate that the associated reasoning
scales up smoothly. In the other direction, we retain the simple
language but enhance the complexity of the properties we prove.

In the examples we present, we will omit the outermost universal
quantifiers when we write specification formulas,
using the convention that tokens 
given by capital letters denote variables that are implicitly
universally quantified over the entire formula. We will also assume
the availability of two special predicates: the binary infix predicate
$=$ for each type that is defined by the clause $X = X \triangleq
\top$ and, for each nominal type, the unary predicate 
{\sl name} that is defined by the clause
$(\nabla x. \name x) \triangleq \top.$ 
Finally, we will assume that the formula 
\begin{equation*}
\forall L, E.\nabla x.~\member {(E\ x)} L \supset \exists E'.~ E =
\lambda y.E',
\end{equation*}
is derivable. This formula, which can be proved by a straightforward
induction on the definition of {\sl member}, states that if a list
does not contain a nominal constant then no element of the list can
contain that constant. 

We will leave out many details of proofs in our presentation,
restricting ourselves to indicating the general structure of the
argument and to highlighting especially interesting applications of
inference rules and the use of induction.

\subsection{Type Preservation for the Simply Typed $\lambda$-Calculus}

\begin{figure}
\begin{align*}
&\eval {(\lam A R)} (\lam A R) \\
&\eval M (\lam A R) \supset \eval {(R\ N)} V \supset \eval {(\app
    M N)} V\\[6pt]
&\of M (\arr A B) \supset \of N A \supset \of {(\app M N)} B \\
&\forall x.(\of x A \supset \of {(R\ x)} B) \supset \of{(\lam A
    R)}{(\arr A B)}
\end{align*}
\caption{Evaluation and typing in the simply typed $\lambda$-calculus}
\label{fig:stlc}
\end{figure}

We recall the encoding of the simply typed $\lambda$-calculus in \hh that was
presented in Section~\ref{ssec:two-level-example}.
We use $ty$ and $tm$ as
the types for \hh terms that encode the types and terms of the
(object) $\lambda$-calculus. The \hh constants
$i : ty$ and $\hsl{arr} : ty \to ty \to ty$ are used to denote a base
type and the arrow type; we
assume for simplicity that there is only one base type in the
object language. The \hh constants $app : tm \to tm \to tm$ and $lam :
ty \to (tm \to tm) \to tm$ are used to denote object-level
applications and (typed) abstractions. In this context, call-by-name
evaluation and 
(monomorphic) typing for the simply typed $\lambda$-calculus can be
specified by the \hh formulas as shown in Figure~\ref{fig:stlc}.

Consider now proving that evaluation in the simply typed 
$\lambda$-calculus preserves typing. Stated in terms
of the encoding in \hh, this property can be expressed through the
following formula in \logic:
\begin{equation}\label{form:type-pres}
\forall E, V, A.~ \{\eval E V\} \supset \{\of E A\} \supset \{\of V A\}.
\end{equation}
We show below how a proof can be constructed in Abella of a sequent with
only this formula on the right.

Using the right rules for the universal quantifier and implication,
the starting goal can be reduced to the subgoal corresponding to the
sequent 
\begin{equation*}
\{\eval E V\}, \{\of E A\} \lra \{\of V A\}.
\end{equation*}
We can prove this sequent by induction on $\{\eval E V\}$ using
the rest of the sequent to generate the induction
invariant. Let us abbreviate that induction hypothesis, namely,
$[\forall E, V, A.~ \{\eval E V\}^* \supset \{\of E A\} \supset \{\of
  V A\}]$ by $\IH$.
The resulting induction yields two sequents, one for each clause
defining {\sl eval}.  The base case, namely,
\begin{equation*}
\IH, \{\of {(\lam B R)} A\} \lra \{\of {(\lam B R)} A\}
\end{equation*}
is trivial.  The other case is given by the sequent
\begin{equation*}
\IH, \{\eval M (\lam B R)\}^*, \{\eval {(R\ N)} V\}^*,
\{\of {(\app M N)} A\} \lra \{\of V A\}.
\end{equation*}
Applying case analysis to the typing judgment on the left
yields the sequent
\begin{align*}
\IH, \{\eval M (\lam B R)\}^*, \{\eval {(R\ N)} V\}^*,\\
\{\of M (\arr C A)\}, \{\of N C\} &\lra \{\of V A\}.
\end{align*}
Applying the induction hypothesis to the evaluation and
typing judgments on $M$ yields the sequent
\begin{equation*}
\IH, \ldots, \{\eval {(R\ N)} V\}^*, \{\of N C\}, \{\of {(\lam B
R)} (\arr C A)\} \lra \{\of V A\}.
\end{equation*}
Case analysis can be applied to the new typing judgment and this yields
\begin{equation*}
\IH, \ldots, \{\eval {(R\ N)} V\}^*, \{\of N B\}, \{\of c B \vdash
\of {(R\ c)} A\} \lra \{\of V A\}.
\end{equation*}
Notice that this analysis has forced $B = C$ and thus all
instances of $C$ have been replaced. In the last hypothesis of
this sequent, $c$ is a nominal constant so we can apply the
instantiation property of \hh to obtain $\{\of N B \vdash \of {(R\
  N)} A\}$. We can then use the cut property with the assumption
$\{\of N B\}$ to produce the following sequent.
\begin{equation*}
\IH, \ldots, \{\eval {(R\ N)} V\}^*, \{\of {(R\ N)} A\} \lra \{\of V A\}.
\end{equation*}
Applying the induction hypothesis to the two assumptions displayed
above completes this proof.

Proofs of properties such as the one above involve what is often
called a ``substitution lemma.'' In this case, assuming a conventional
syntax representation, such a lemma would be stated as ``if $B$ has type
$\alpha$ and the variable $x$ and term $t$ have the same type $\beta$,
then $B[t/x]$ has type $\alpha$.''  Such a lemma can be proved using
an induction on the details of the construction of terms and their
binding structure.  Notice that in the proof above, this substitution
lemma comes {\em for free}: it is a direct application of the
cut-admissibility result for \hh.  Of course, the proof of
cut-admissibility requires a detailed induction on the structure of \hh
proofs. As this example illustrates, however, once cut-admissibility has
been established, one should be able to get most substitution lemmas
for free by using such meta-level properties of \hh.

Our ultimate objective is to show the type preservation property for the
simply typed $\lambda$-calculus. We obtain this result from the
property stated in formula~(\ref{form:type-pres}) by using the adequacy
of our encodings. Suppose that $e$ 
evaluates to $v$ and that $\vdash e : a$ holds. Let $\Delta$ be
the clauses in Figure~\ref{fig:stlc}. By the adequacy of these
clauses, which can be proved as shown in
Section~\ref{sec:two-level-reasoning}, we know that $\Delta \vdash
\eval {\phi(e)} {\phi(v)}$ and $\Delta \vdash \of {\phi(e)}
{\phi(a)}$ must have derivations in \hh. Then from the adequacy of
the {\sl seq} encoding of \hh into \logic we know that $\{\eval
{\phi(e)} {\phi(v)}\}$ and $\{\of {\phi(e)} {\phi(a)}\}$ must both
have proofs in \logic. Using the proofs of these two formulas
together with the proof of formula~(\ref{form:type-pres}),
we can construct a proof of
$\{\of {\phi(v)} {\phi(a)}\}$. Then by the adequacy of {\sl seq}, it
must be that $\Delta \vdash \of {\phi(v)} {\phi(a)}$ has a
derivation in \hh. Finally by the adequacy of the clauses in
$\Delta$ it must be that $\vdash v : a$ holds. Notice that one
must prove adequacy for the clauses which make up a specification,
but one does not need to ever re-prove the adequacy of {\sl seq}.
Thus, the two-level logic approach to reasoning does not introduce
any recurring costs with respect to adequacy of the associated
reasoning.

\subsection{Type Uniqueness for the Simply Typed $\lambda$-Calculus}

Proving the formula $[\forall E, A, B.~\{\of E A\} \supset \{\of E B\}
  \supset A = B]$, that is, that types are unique for the simply typed
$\lambda$-calculus, brings out another
important aspect of the two-level logic approach to reasoning: the
reasoning logic can be used to make explicit, and thereby to exploit in
reasoning, properties of terms that arise dynamically when the
specification logic is used to ``carry out'' computations described in
it. Specifically, in this example we will use \logic to characterize
the typing contexts that are constructed in \hh when using
hypothetical judgments to assign types
to abstractions.

In order to prove the theorem about uniqueness of types, we will need
to generalize it to allow for the assignment of types relative to
typing contexts. These typing contexts can be characterized in \logic
by a variant of the {\sl cntx} predicate that we saw in
Section~\ref{sec:logic} that is defined by the following clauses:
\begin{align*}
\ctx \nil \mueq \top && (\nabla x.~ \ctx (\of x A :: L)) \mueq \ctx L.
\end{align*}
It is easy to see that if the judgment $\ctx L$ holds, then $L$ must
be a list of elements  of the form $(\of x A)$ where each $x$ is a
nominal constant that does not appear later in the list. Thus, the
type assignments in $L$ must be to nominal constants and the
assignment to each such constant must be unique.  These properties,
which are needed for proving the uniqueness of typing, 
are written as the following formulas in \logic:
\begin{equation}
\label{eq:ctx_member}
\forall X, A, L.~ \ctx L \supset \member {(\of X A)} L \supset
\name X
\end{equation}
\begin{equation}
\label{eq:ctx_unique}
\forall X, A, B, L.~ \ctx L \supset \member {(\of X A)} L \supset
  \member {(\of X B)} L \supset A = B.
\end{equation}
Both formulas can be established as lemmas in \logic by a simple induction
on the structure of the {\sl ctx} definition. 
Notice that in the second formula, the universally quantifier over
$X$ could have been replaced by the generic quantifier over $X$.
We also note that the proof of this second formula makes use of the
general lemma about list membership and nominal constants described at
the beginning of this section.

The generalization of the type uniqueness theorem is now given as the
following formula:
\begin{equation*}
\forall E, A, B, L.~ \ctx L \supset \{L \vdash \of E A\} \supset
\{L \vdash \of E B\} \supset A = B.
\end{equation*}
Attempting to prove this formula yields the sequent
\begin{equation*}
\ctx L, \{L \vdash \of E A\}, \{L \vdash \of E B\} \lra A = B.
\end{equation*}
Applying induction on the first typing judgment with the following
inductive
hypothesis (again denoted by \IH)
\begin{equation*}
\forall E, A, B, L.~ \ctx L \supset \{L \vdash \of E A\}^* \supset
\{L \vdash \of E B\} \supset A = B.
\end{equation*}
results in three cases. The first case is
\begin{equation*}
\IH, \ctx L, \member {(\of E A)} L, \{L \vdash \of E B\} \lra A = B.
\end{equation*}
We can apply lemma~(\ref{eq:ctx_member}) here to obtain 
\begin{equation*}
\IH, \ctx L, \member {(\of E A)} L, \name E, \{L \vdash \of E B\}
\lra A = B.
\end{equation*}
Applying case analysis to the assumption $\name E$ 
leads to a single premise since there is a most general upper sequent
for this use of $\defL$.
In particular, $E$ is replaced by a nominal constant $c$
and every other variable is raised over this constant. Thus we
have the following sequent:
\begin{equation*}
\IH, \ctx (L\ c), \member {(\of c (A\ c))} (L\ c), \{L\ c \vdash
\of c (B\ c)\} \lra A\ c = B\ c.
\end{equation*}
Now case analysis on the remaining typing assumption results in the 
single sequent
\begin{equation*}
\IH, \ctx (L\ c), \member {(\of c (A\ c))} (L\ c), \member {(\of c
  (B\ c))} (L\ c) \lra A\ c = B\ c.
\end{equation*}
At this point we can apply lemma~(\ref{eq:ctx_unique}) to finish
this case.

The second of the three original cases is the sequent
\begin{equation*}
\IH, \ctx L, \{L \vdash \of M (\arr C A)\}^*, \{L \vdash \of N C\}^*, \{L
\vdash \of {(\app M N)} B\} \lra A = B.
\end{equation*}
Now we can perform case analysis on the remaining typing
assumption for $\app M N$. This results in two cases. The first is
that $\of {(\app M N)} B$ may occur in the list $L$. This case can
be handled using lemma~(\ref{eq:ctx_member}), \ie, we can
determine $\name {(\app M N)}$ which when subjected to case
analysis will result in zero cases (that is, it is recognized as
a false assumption). The other case is 
\begin{equation*}
\IH, \ctx L, \{L \vdash \of M (\arr C A)\}^*, \ldots, \{L \vdash \of M (\arr D
B)\}, \ldots \lra A = B.
\end{equation*}
At this point we can apply the induction hypothesis to the two
typing judgments for $M$ to determine that $\arr C A = \arr D B$
and therefore $A = B$.

The remaining case in the original proof is the sequent
\begin{equation*}
\IH, \ctx L, \{L, \of c C \vdash \of {(R\ c)} D\}^*, \{L \vdash \of
{(\lam C R)} B\} \lra \arr C D = B.
\end{equation*}
Here $c$ is a nominal constant. Case analysis on the typing
judgment for $\lam C R$ results in two cases. Again, the first one
can be dismissed using lemma~(\ref{eq:ctx_member}). The second one
is as follows.
\begin{equation*}
\IH, \ctx L, \{L, \of c C \vdash \of {(R\ c)} D\}^*, \{L, \of c C
\vdash \of {(R\ c)} F\} \lra \arr C D = \arr C F.
\end{equation*}
Here we have opted to use the nominal constant $c$ in deconstructing
this second typing judgment. Any other choice is equally valid and
does not affect the proof.  In order to use
the induction hypothesis we must be able to show that $\ctx (\of c C
:: L)$ holds: but this is immediate from the definition of {\sl ctx}
and the fact that $c$ is a nominal constant which does not appear in
$L$. Therefore we can use the induction hypothesis and determine
that $D = F$, thus finishing the proof.

\subsection{Extension to the Language of PCF}

We now extend the specification of the simply typed $\lambda$-calculus
to treat an abstract version of the programming language PCF presented 
by \cite{plotkin77}.  To do this,
we replace the base type $i : ty$ with the types for numbers 
$\hsl{num} : ty$ and booleans $\hsl{bool} : ty$.  We also enrich the set of
terms by allowing the following constants.
\begin{align*}
\hsl{zero} &: tm & \hsl{succ} &: tm \to tm & \hsl{if} &: tm \to tm \to tm \to tm \\
\hsl{true} &: tm & \hsl{pred} &: tm \to tm & \hsl{rec} &: ty \to (tm \to tm) \to tm \\
\hsl{false} &: tm & \hsl{iszero} &: tm \to tm &
\end{align*}
Using these, the specification for evaluation and typing in PCF
is presented in Figure~\ref{fig:pcf}.

\begin{figure}
\begin{align*}
&\eval \zerop \zerop \\
&\eval \truep \truep \\
&\eval \falsep \falsep \\
&\eval M V \supset \eval {(\succp M)} {(\succp V)} \\
&\eval M \zerop \supset \eval {(\predp M)} \zerop \\
&\eval M (\succp V) \supset \eval {(\predp M)} V \\
&\eval M \zerop \supset \eval {(\iszerop M)} \truep \\
&\eval M (\succp V) \supset \eval {(\iszerop M)} \falsep \\
&\eval M \truep \supset \eval {N_1} V \supset \eval {(\ifp M {N_1}
    {N_2})} V \\
&\eval M \falsep \supset \eval {N_2} V \supset \eval {(\ifp M
    {N_1} {N_2})} V \\
&\eval {(\lam A R)} (\lam A R) \\
&\eval M (\lam A R) \supset \eval {(R\ N)} V \supset \eval {(\app
    M N)} V \\
&\eval {(R\ (\recp A R))} V \supset \eval {(\recp A R)} V \\
& \\
&\of \zerop \nump \\
&\of \truep \boolp \\
&\of \falsep \boolp \\
&\of M \nump \supset \of {(\succp M)} \nump \\
&\of M \nump \supset \of {(\predp M)} \nump \\
&\of M \nump \supset \of {(\iszerop M)} \boolp \\
&\of M \boolp \supset \of {N_1} A \supset \of {N_2} A \supset \of
  {(\ifp M {N_1} {N_2})} A \\
&\of M (\arr A B) \supset \of N A \supset \of {(\app M N)} B \\
&(\forall x. \of x A \supset \of {(R\ x)} B) \supset \of{(\lam A
    R)}{(\arr A B)} \\
&(\forall x. \of x A \supset \of {(R\ x)} A) \supset \of {(\recp A
    R)} A
\end{align*}
\caption{Evaluation and typing in PCF}
\label{fig:pcf}
\end{figure}

We shall not repeat the proofs of type preservation and type
uniqueness for PCF, but rather we will explain how these proofs
differ from the ones for the simply typed $\lambda$-calculus.
First, for type preservation, the statement is unchanged:
\begin{equation*}
\forall E, V, A.~ \{\eval E V\} \supset \{\of E A\} \supset \{\of V A\}.
\end{equation*}
The basic structure of this proof is the same, however, when we
induct on $\{\eval E V\}$ we get 13 cases instead of two, since {\sl
  eval\/} has that many more cases now.
These additional cases are either easy or similar
to the cases in the earlier version of the proof. The substitution
property for typing judgments is again obtained for free using the
instantiation and cut properties of \hh.  The only increase in
proof size is due to a widening of the central case analysis.
The story for type uniqueness is the same: 
since typing contexts have not been changed, the definition of {\sl
ctx} is as before and the proof of the formula
\begin{equation*}
\forall E, A, B, L.~ \ctx L \supset \{L \vdash \of E A\} \supset
\{L \vdash \of E B\} \supset A = B
\end{equation*}
proceeds as before but with additional cases as expected from the
additional clauses in the specification of typing.

\subsection{Comparing Paths in $\lambda$-Terms}
\label{sec:path-equiv-lambda}

\begin{figure}[t]
\begin{align*}
&\term M \supset \term N \supset \term (\app M N) \\
&(\forall x. \term x \supset \term (R\ x)) \supset \term (\uabs R) \\[6pt]
&\pathp M \pdone \\
&\pathp M P \supset \pathp {(\app M N)} {(\pleft P)} \\
&\pathp N P \supset \pathp {(\app M N)} {(\pright P)} \\
&(\forall x.\forall p. \pathp x p \supset \pathp {(R\ x)} {(S\ p)})
\supset \pathp {(\uabs R)} {(\bnd S)}
\end{align*}
\caption{Specification of paths through $\lambda$-terms}
\label{fig:spec-path}
\end{figure}

Terms in the untyped, pure $\lambda$-calculus can be visualized as
tree structures. As such, we can define paths in a term as paths  that
start at the root in the corresponding tree.  We shall formally prove
here that if every path in one $\lambda$-term is also a path in
another $\lambda$-term, then the two terms are equal.

To formalize this theorem, we first need a representation of untyped
$\lambda$-terms and paths in \hh.  We introduce the two types {\sl tm}
and {\sl pt} for this purpose and we use the constructors shown below.
\begin{align*}
\hsl{app}  &: tm \to tm \to tm &
\hsl{done} &: pt &
\hsl{left} &: pt \to pt \\
\hsl{abs}  &: (tm \to tm) \to tm & 
\hsl{bnd} &: (pt \to pt) \to pt  &
\hsl{right} &: pt \to pt
\end{align*}
Notice that since we are concerned with only pure $\lambda$-terms, we
only need the two constructors {\sl app} and {\sl abs} for
representing them.  

We now introduce the predicates {\sl term} and {\sl path} defined by 
the specification logic formulas in Figure~\ref{fig:spec-path}.  Note
that we allow partial paths using \pdone.  Notice also that the
\logic formula 
\[
\forall R, S.
\{\pathp {(\uabs R)}{(\bnd S)}\}\supset 
(\forall x.\forall p. \{\pathp x p\}
      \supset \{\pathp {(R\ x)} {(S\ p)}\})
\]
is a kind of converse to the last clause specifying {\sl path} and is
also trivial to prove.
Thus, if we have a path $(\bnd S)$ through the
term $(\uabs R)$ and a path $P$ through term $N$, then the result of
substituting path $P$ into the abstraction $S$ is a path in the term
resulting from substituting $N$ into the abstraction $R$.  This
formula is another example of  a ``substitution lemma for free.''

We wish now to prove the theorem:
\begin{align}\label{prop:path-equiv}
&\forall M, N.~ \{\term M\} \supset (\forall P.~ \{\pathp M P\}
\supset \{\pathp N P\}) \supset M = N.
\end{align}
Since induction in \logic is an introduction rule for defined
predicates, the assumption $\{\term M\}$ is placed in this formula to
enable induction on the structure of $M$.
Before we prove this formula, we need to strengthen it. In particular,
when $M$ is an abstraction we 
need to consider how the contexts for the {\sl term} and {\sl path}
judgments will grow.  The defined predicate
{\sl ctxs} describes how these two contexts are related.
\begin{align*}
\ctxs {\nil} {\nil} \mueq \top && (\nabla x. \nabla p. \ctxs {(\term x
  :: L)} {(\pathp x p :: K)}) \mueq \ctxs L K.
\end{align*}
Along with this definition, we need the following lemmas which allow
us to extract information about a term based on its membership in one
of the contexts described by {\sl ctxs}.
\begin{align*}
&\forall X, L, K.~ \ctxs L K \supset \member {(\term X)} L \supset
\name X \land \exists P.~\member {(\pathp X P)} K \\
& \forall X, P, L, K.~ \ctxs L K \supset \member {(\pathp X P)} K
\supset \name X \land \name P.
\end{align*}
The proofs of both lemmas are by straightforward induction on the {\sl
  member} hypotheses.

We can state the strengthened version of the theorem as the following
lemma. 
\begin{multline*}
\forall L, K, M, N.~ \ctxs L K \supset \{L \vdash \term M\} \supset \\
(\forall P.~ \{K \vdash \pathp M P\} \supset \{K \vdash
\pathp N P\}) \supset M = N.
\end{multline*}
The proof of this lemma proceeds by induction on $\{L \vdash \term
M\}$.  The base case needs the following lemma, which is proved by
induction on one of the {\sl member} hypotheses and which uses the
general lemma about list membership and nominal constants described in
the preamble of this section.
\begin{multline*}
\forall L, K, X_1, X_2, P.~ \ctxs L K \supset
\member {(\pathp {X_1} P)} K \supset \\ \member {(\pathp
  {X_2} P)} K \supset X_1 = X_2.
\end{multline*}
In the other cases of the proof, we need to show that the
top-level constructor of $M$ is also the top-level constructor of
$N$. We do this by constructing a partial path through the
top-level constructor of $M$: since paths in $M$ are also paths in
$N$, the top-level constructor of $N$ must match that of $M$.
Once we know that the top-level
constructors are the same, we can use the assumption that all paths in
$M$ are paths in $N$ to show that all paths in an immediate
subterm of $M$ are paths in the corresponding immediate subterm of
$N$. Then by induction we can conclude that those subterms are
equal.

There is one technical complication in the proof of path
equivalence which comes from the inductive case concerning
abstractions. Suppose $M = \uabs R$ and $N = \uabs R'$. Here we
know
\begin{equation*}
\forall P.~ \{K \vdash \pathp {(\uabs R)} P\} \supset
\{K \vdash \pathp {(\uabs R')} P\}
\end{equation*}
but in order to use the inductive hypothesis we must show
\begin{equation*}
\forall P.~ \{K, \pathp x p \vdash \pathp {(R\ x)} P\} \supset
\{K, \pathp x p \vdash \pathp {(R'\ x)} P\},
\end{equation*}
where $x$ and $p$ are nominal constants. Now the problem is that when
we go to prove this latter formula, the $\forallR$ rule says that we
must replace $P$ by $(P'\ p\ x)$ for some new eigenvariable $P'$. Note
that $P'$ is raised over both $p$ and $x$ even though the dependency
on $x$ must be vacuous.
The following lemma establishes this vacuity and is used to finish
this case of the proof.
\begin{align*}
&\forall K, M, P. \nabla x, p.~ \{K, \pathp x p \vdash \pathp {(M\ x)}
{(P\ p\ x)}\} \supset \exists P'.~ P = \lambda z.P'
\end{align*}
This lemma is proved by induction on the {\sl path} judgment and uses the
general lemma about nominal constants and list membership. Note that we
single out $\pathp x p$ being the first member of the context even
though new {\sl path} assumptions may be added during induction. This
is not a problem since we can always use the property of \hh which
allows contexts to be freely rearranged.  With this issue resolved,
the proof of this theorem can now be completed.

In the theorem about paths, we have encoded the the property that all
paths in $m$ are paths in $n$ via the formula 
\begin{equation}
\label{form:path}
\forall P.~ \{\pathp {\phi(m)} P\} \supset \{\pathp {\phi(n)} P\}.
\end{equation}
There is a question about the adequacy of this encoding even after we
have established the adequacy of our representation of terms and
paths and we have shown that, 
%% This theorem about paths has a certain richness in its
%% statement which may raise questions about adequacy.
%% In particular, given two $\lambda$-terms $m$ and $n$, we
%% encode the property that all paths in $m$ are paths in $n$ via the
%% formula
%% \begin{equation}
%% \label{form:path}
%% \forall P.~ \{\pathp {\phi(m)} P\} \supset \{\pathp {\phi(n)} P\}.
%% \end{equation}
%% To show that this formula is an adequate encoding of the
%% stated property, assume that we already know that our representation
%% of terms and paths is adequate and that
for all terms $m$ and paths $p$, $m$ has path $p$ if and only if $\{\pathp
{\phi(m)} {\phi(p)}\}$ is provable in \logic. To resolve this question
in one direction, assume that the formula~(\ref{form:path}) is
provable. Let $p$ be any path in $m$ so that we have a proof of
$\{\pathp {\phi(m)} {\phi(p)}\}$. Using formula~(\ref{form:path})
and this proof we can construct a proof of $\{\pathp {\phi(n)}
{\phi(p)}\}$. Thus $n$ has path $p$. For the 
other direction, we argue that if every path in $m$ is a path in $n$ then
we can prove formula~(\ref{form:path}). Such a proof reduces to
constructing a derivation of the sequent
\begin{equation}
\{\pathp {\phi(m)} P\} \lra \{\pathp {\phi(n)} P\}.
\end{equation}
We can construct a proof of this sequent by repeatedly unfolding
$\{\pathp {\phi(m)} P\}$ and the new hypotheses which result from
it. This process will terminate since $\phi(m)$ is a finite term
with no variables and the recursive clauses of {\sl path} always
deconstruct their first argument. The sequents which result from
this repeated case analysis will have the form $\lra \{\pathp
{\phi(n)} P\}$ for some instance of $P$ such that $\lra \{\pathp
{\phi(m)} P\}$ is provable. By the assumption of adequacy for the {\sl
path} predicate, we know $P = \phi(p)$ where $p$ is a path in $m$.
Thus $p$ is also a path in $n$ and thus each sequent $\lra
\{\pathp {\phi(n)} P\}$ is provable.

\subsection{Other Examples}
\label{sec:other-examples}

There are many other examples of topics that have been completely
formalized within \logic and checked using the Abella prover.  We list
some of these examples here: complete details of the proofs can be
found on the website for Abella \citep{abella.website}.

\begin{figure}[t]
\begin{align*}
& \step {(\app {(\lam A R)} M)} {(R\ M)} \\
& \step M {M'} \supset \step {(\app M N)} {(\app {M'} N)} \\
& \step N {N'} \supset \step {(\app M N)} {(\app M {N'})} \\
& (\forall x. \step {(R\ x)} {(R'\ x)}) \supset \step {(\lam A R)}
{(\lam A R')}
\end{align*}
\caption{One step $\beta$-reduction in the simply typed $\lambda$-calculus}
\label{fig:step}
\end{figure}

\paragraph{Meta-theory of the $\lambda$-calculus}

We have used Abella to specify both big-step and small-step evaluation
for untyped $\lambda$-terms and then to prove that they compute
the same values and that they are both determinate and 
type-preserving.  We have also encoded a proof of the Church-Rosser
theorem and have also proved strong normalization for the simply typed
$\lambda$-calculus.  The latter theorem and proof deserve a few
additional words.    Strong normalization for the
$\lambda$-calculus can be defined elegantly as
\begin{equation*}
\sn M \mueq \forall N. \{\step M N\} \supset \sn N,
\end{equation*}
where {\sl step} (specified in Figure~\ref{fig:step}) relates two
terms when the second is the replacement of exactly one $\beta$-redex
in the first.
Induction on {\sl sn} corresponds to induction on the tree of possible
$\beta$-reductions for a term which in this case can be used in place
of induction on the longest possible length of a $\beta$-reduction.
Using the predicate {\sl of} defined in
Figure~\ref{fig:stlc}, the strong normalization theorem for the simply
typed $\lambda$-calculus is stated simply as
\begin{equation*}
\forall M, A. \{\of M A\} \supset \sn M.
\end{equation*}
The proof of this theorem uses a logical relations style argument
based on the predicate {\sl reduce} defined as
\begin{align*}
& \reduce M i \mueq \{\of M i\} \land \sn M \\
& \reduce M {(\arr A B)} \mueq \{\of M {(\arr A B)}\} \land \forall U.~
\reduce U A \supset \reduce {(\app M U)} B.
\end{align*}
Abella allows such a definition although it does not satisfy the
stratification condition described in Section~\ref{sec:logic}.  As we
mention in Section~\ref{sec:conclusion}, more flexible notions of
stratification need to be identified and validated in order to justify
this proof.

\paragraph{Meta-theory of the $\pi$-calculus}

We have specified the semantics of the finite $\pi$-calculus using the
specification logic and formalized the notion of open bisimulation
using a co-inductive definition in the reasoning logic. We have shown that
open bisimulation is an equivalence relation and a congruence using
this formalization. This formalization constitutes an elegant treatment of the
$\pi$-calculus where all issues involving bindings, names, and
substitutions are handled declaratively without explicit
side-conditions.

\paragraph{The POPLmark challenge problems}

The POPLmark challenge \citep{aydemir05tphols} is a selection of
problems which highlights the traditional difficulties in reasoning
about systems which manipulate objects with binding. The particular
tasks of the challenge involve reasoning about evaluation, typing, and
subtyping for $F_{<:}$, a $\lambda$-calculus with bounded subtype
polymorphism. We have solved parts 1a and 2a of this challenge using
Abella, which represent the fundamental reasoning tasks involving
objects with binding.

\paragraph{Cut-elimination}

We have shown that the cut rule can be eliminated from LJ while
preserving the provability relation. The encoding of sequents in
our specification logic used hypothetical judgments to represent LJ
hypotheses and generic judgments to represent LJ universals. This
allowed the cut-elimination proof to take advantage of Abella's
built-in treatment of meta-properties of the specification logic.

%%% Local Variables: 
%%% mode: latex
%%% TeX-master: "root"
%%% End: 

% LocalWords:  PCF tm ty sequents eval ctx subgoal num bool succ pred iszero LJ
% LocalWords:  bnd versa ctxs subterms eigenvariable monomorphic booleans unary
% LocalWords:  subterm Abella Rosser disjointly bisimulation declaratively cntx
% LocalWords:  POPLmark subtyping Abella's plotkin abella aydemir tphols sn

\section{Related Work}
\label{sec:related-work}

The range of applications that we have demonstrated for our reasoning
logic \logic depends on its strong declarative treatment of binding 
as well as its treatment of fixed points ({\em i.e.}, induction
and co-induction). In comparing our work to the many other 
research efforts devoted to building theorem provers that can
reason about specifications of computations, it is convenient to 
characterize the latter approaches using these two axes of logical
expressiveness. Some of these systems 
start with a clean and comprehensive foundation for fixed points and
(co)induction, treating the notion of 
of binding as something that can be implemented later within such an
inductive logic. 
Other systems start with a logically supported approach to binding and
then later provide some aspects of inductive reasoning over binding
structure.  
We use this coarse classification below to organize our comments about
related efforts.

\subsection{Inductive Frameworks with Treatments of Binding Added}

Many systems for reasoning about computations start with established
inductive logic theorem provers such as Coq \citep{bertot04book}
(based on the {\em Calculus of Inductive Constructions}
\citep{coquand88colog}) and Isabelle/HOL \citep{nipkow02book}, and
then use those systems to build approaches to binding and
substitution.
% DM: dropped the reference to chargueraud09ln.  Seems that this is
% really just a technique based on de Bruijn's ideas:  it's
% mechanization is described in aydemir08popl.
We discuss three examples of this approach: the locally nameless
representation, the Nominal package for
Isabelle/HOL \citep{urban08jar}, and Hybrid \citep{felty10jar}.

% Locally Nameless

The locally nameless representation of binding structure uses de
Bruijn indices for bound variables and names for free 
variables.  The central benefits of this approach are that
$\alpha$-equivalent terms are syntactically equal, the statements
of lemmas and theorems rarely need to talk about arithmetical
operations over de Bruijn indices, and capture-avoiding
substitution can be defined in a straightforward and structurally
recursive way.  However, one must still define this substitution 
manually and prove lemmas about its behavior.  Additionally, there 
is no device like $\nabla$ for quantifying over fresh variable
names. Instead, practitioners of the locally nameless approach (see, for
example, \cite{aydemir08popl}) advocate an
encoding of such quantification using {\em cofinite
quantification}, \ie, quantification over all names not belonging
to some arbitrary, finite set.  This technique, however, 
% is a
% DM:  Andrew, I like this term ``leaky abstraction'' but the example
% you gave here about free variable renaming doesn't seem a convincing
% example.  Can names/bindings escape their intended scopes?
% ``leaky abstraction'' in that one must still occasionally prove, for
% example, 
still requires sometimes explicitly proving 
that free variables can be
renamed while preserving provability of a judgment.

% Nominal package

The Nominal package for Isabelle/HOL automates the formalization
of alpha-equivalence classes based on ideas from nominal logic
\citep{pitts03ic}. The user is then left to define and
reason about a notion of capture-avoiding substitution over terms
constructed with such alpha-equivalence classes. Reasoning over
open terms is supported in the Nominal package via the nominal
logic $\new$-quantifier which has similarities to the
$\nabla$-quantifier.
However, the $\new$-quantifier is ``over-worked'' in
the nominal approach since it is also used to introduce names
which are bound by name abstractions. This creates some additional
difficulties such as when introducing functions and predicates in
the nominal approach one must prove properties which state that
name swapping does not change the results of a function or the
provability of a predicate---a property which is enforceable
statically for definitions of predicates in \logic due to the
separation between free and bound variables.

% Hybrid

Hybrid adds support to traditional theorem provers such as Coq and
Isabelle/HOL for reasoning about binding structures by
translating such structure into a de Bruijn representation.
The logic of the theorem prover
then serves as the meta-logic in which reasoning is conducted. This
approach necessarily produces more overhead during reasoning due to
the occasional need to reason about the effects of the translation,
although one might expect that such reasoning can eventually be automated.
Hybrid is often used in a two-level
logic approach using a specification logic that is essentially the
same \hh specification language considered in this paper.
The Hybrid system, by design, lacks a reasoning logic with a device like
the $\nabla$-quantifier for reasoning about open terms and generic
judgments. 
Recent work has suggested that such a device is not
necessary for simple reasoning tasks such as type uniqueness
arguments \citep{felty09ppdp}, although it is unclear if the Hybrid
approach will scale to problems such as those
proposed by the POPLmark challenge \citep{aydemir05tphols}.  For
these types of problems one needs to recognize as equivalent those
judgments which differ only in the renaming of free variables.
Such a property is built into \logic through the use of nominal
constants to denote such free variables. To use such an approach in
Hybrid, one will have to manually develop and prove properties about
notions of variable permutations.

\subsection{Binding Frameworks with Treatments of Induction Added}

There are a variety of systems for reasoning about computations which
take binding as a primitive notion and then attempt to define
separately notions of induction over that structure.  Many of these
start with the LF logical framework \citep{harper93jacm}, a
dependently typed $\lambda$-calculus with a direct treatment of
variable binding.  While the LF type system can be used to describe
both the structure and behavior of many computational systems, it does
not include a notion of induction: inductive arguments about LF
specifications are typically supported by constructing a second layer
on top of LF.

Twelf \citep{pfenning99cade}, the most popular tool for reasoning
about LF specifications, provides an operational semantics for LF that 
defines recursive relations over LF terms.  Subject
to some side-conditions, these relations can then be interpreted as
proofs about LF specifications.  Similar functional approaches have
been developed starting with ${\cal M}_2^+$ \citep{schurmann00phd}, a
simple meta-logic for reasoning over LF representations where proof
terms are represented as recursive functions.  More recent work
includes the Delphin \citep{poswolsky08lfmtp} and Beluga
\citep{pientka08popl} functional languages which can be used in the
same spirit as ${\cal M}_2^+$. New work by \cite{licata08lics}
proposes a language which combines LF with recursive functions over LF
so that a strict separation into levels is no longer needed. In all of
these approaches, however, side-conditions for termination and
coverage are required and algorithms have been devised to check 
for such properties.  Since termination and coverage are in general 
undecidable, such algorithms are necessarily incomplete.

\subsection{The Development of a Logic for both Bindings and Fixed
  Points} 

The logic \logic is the result of an extended effort to design a
single logic that integrates induction and co-induction with the
ability to reason flexibly about bindings.  The $\lambda$Prolog language
\citep{nadathur88iclp} provided a starting point as a 
specification language 
%GN This makes it sound as if \hh came first and lp extended it. It
%also sounds like an unnecessary mention, given that we have spoken
%about the connection a few times already.
%(extending \hh) 
that allowed a completely
declarative treatment of binding.  In order to support reasoning about 
specifications written in the \hh subset of $\lambda$Prolog,
\cite{mcdowell00tcs} developed the 
two-level logic approach used in this paper but with a much weaker
reasoning logic called \FOLDN.  That logic provided induction on natural
numbers but did not contain $\nabla$-quantification.  As a result of
this missing ingredient, reasoning about object-level bindings became
unduly complicated; see, for example, the discussion on explicit
eigenvariable encoding in \citep{mcdowell00tcs}.

The $\nabla$-quantifier was first introduced in
\citep{miller05tocl}. The logic that was first proposed did not
include inference rules for induction and 
co-induction but these were added shortly thereafter by \cite{tiu04phd}. The
initial logics adopted a minimalistic view of the $\nabla$-quantifier
that turned out to be inadequate for many instances of inductive
reasoning over binding structures. To redress this situation
\cite{tiu06lfmtp} proposed the addition of the $\nabla$-exchange and
$\nabla$-strengthening rules and developed the nominal constant based
treatment of the $\nabla$-quantifier used in this paper. The resulting
logic still did not have the ability to concisely characterize
occurrences of nominal constants in terms and was consequently awkward
to use in reasoning about open terms and contexts. The missing piece
was provided by the notion of nominal abstraction
\cite{gacek09corr}. This final logic, \logic, combines into one 
proof system, the two separate components for reasoning about
fixed points and about binding.  These components are independently
constructed yet their interaction is well-behaved and quite
useful.

%% DM: the following line seem unclear and unnecessary
% The development of all of these logics has focused on proof theory: in
% all cases, all proposed logics were given sequent calculus proofs
% system for which cut-elimination was proved.  Satisfying
% cut-elimination is, indeed, strong evidence that different logical
% features interact well.

%%% Local Variables: 
%%% mode: latex
%%% TeX-master: "root"
%%% End: 

% LocalWords:  Coq HOL logics POPLmark Delphin LF harper jacm urmann's phd cade
% LocalWords:  schurmann pfenning bisimulation CCS cntx subst co datatype ln tr
% LocalWords:  bertot nipkow chargueraud felty Bruijn indices pitts ic ppdp de
% LocalWords:  aydemir tphols Twelf poswolsky lfmtp Beluga pientka popl colog
% LocalWords:  undecidability nadathur iclp eigenvariable mcdowell tcs coquand

% LocalWords:  Prolog

\section{Conclusions and Future Work}
\label{sec:conclusion}

We have presented an intuitionistic logic, \logic, in which binding is
treated directly using the
$\nabla$-quantifier (both in formulas and the head of definitions) and 
in which least and greatest fixed points are
treated directly using inference rules for induction and
co-induction.  In a logic that has this kind of expressiveness, it is
possible to inductively define proofs systems for specification
logics, such as \hh.  This makes it possible to use a theorem-proving
approach in which the specification logic is
used in an intrinsic way and in which reasoning takes place through a
transparent embedding of that logic in 
the richer reasoning logic. We have described a system called
Abella that exploits this two-level logic approach and we have shown
its flexibility and power through a sequence of reasoning
examples. While the illustrations we have been able to consider in
this paper are limited, Abella has had a number of
significant theorem-proving successes that are described more
completely on the web page associated with it. 

Experience with the two-level logic approach to reasoning has provided
us with insights into possible ways to enhance the logic \logic
and the methodology built into Abella. We indicate a few such 
directions that we intend to pursue in the near future.

\paragraph{More permissive stratification conditions for definitions}
The current stratification condition for definitions in \logic is
somewhat simplistic: that condition rules out seemingly well-behaved
definitions such as that of the reducibility relation used in logical
relations arguments; see Section~\ref{sec:other-examples} for
details.
One could imagine a more sophisticated condition
which would allow definitions to be stratified based on an ordering
relation over the arguments of the predicate being defined. The proof
theoretic arguments needed to prove cut-elimination for a logic with
such definitions seem rather delicate, particularly since we allow
substitutions which may interfere with any ordering based on term
structure.

\paragraph{Contexts are special}
In principle, provability in the specification logic is captured by
an inductive definition of the {\sl seq} predicate; in practice, it
has been most useful for Abella to provide some special treatment of
that predicate (via the $\{\cdot\vdash\cdot\}$ notation).  Similarly,
while contexts are, in principle, just another list structure, it
seems likely that they should also have some special attributes
associate to them.  As some examples illustrated, the current practice
requires  stating a definition describing a 
context, proving various inversion lemmas about membership in such
contexts, and then applying these lemmas at the appropriate times.
Treating context as special objects should make it possible to
automate several of these lemmas or to arrange things so that such
lemmas are not needed but have their effects embedded into the
prover. 

\paragraph{Types-as-predicates}
As we have described the logic \logic, there is no direct connection
between predicates (on which we may apply induction) and the simple
types attributed to variables.  The description of the type and
its constructors is not sufficient: it is necessary to define a
predicate that describes the members of the type.   For example, 
if we wish to do induction on the structure of untyped $\lambda$-terms
(as in Section~\ref{sec:path-equiv-lambda}), we need to build the
predicate {\sl term} from the description of the type {\sl tm}.  Linking
simple types to the predicates that define them is a natural
enhancement to a theorem prover for \logic.

\paragraph{Alternate specification logics and proof systems}
In this paper, we fixed the specification logic to be \hh and we fixed
the proof system for \hh to be based on goal-directed proof search.  
Clearly some applications of the two-level logic approach might
benefit from using a different proof system (based on, say, bottom-up
proof search) or a different logic.   For example,
\cite{mcdowell02tocl} showed that switching to a linear logic
specification logic made it possible to treat programming languages
with references.  
More concretely, we have implemented a
full hereditary Harrop formula specification logic in Abella and have
begun experimenting with reasoning over it.

\paragraph{Automating proof search}
Abella currently relies extensively on user guidance in constructing
proofs. Recent work has developed formal theorems
and implementation techniques for
structuring proof search in \logic-like logics: see, for example,
\citep{baelde07cade,baelde08phd}. It would be interesting to use such 
results to build a greater degree of automation into Abella.  
% Dropped the reference to baelde07lpar since it has the same time as
% the baelde09tr and since the latter subsumes the former.  Also
% dropped the reference to the paper with Snow and Viel since it never
% appearred and needs a lot of work.

%%% Local Variables: 
%%% mode: latex
%%% TeX-master: "root"
%%% End: 

% LocalWords:  Harrop Abella chirimar phd mcdowell tocl baelde lpar seq tm cade
% LocalWords:  intuitionistic

\section{Acknowledgements}
\label{sec:ack}

This work
has been supported by the National Science Foundation grants
CCR-0429572 and CCF-0917140 and by INRIA through the
``Equipes Associ{\'e}es'' Slimmer. Opinions, findings, and conclusions
or recommendations expressed in this papers are those of the authors
and do not necessarily reflect the views of the National Science
Foundation.

%%% Local Variables:
%%% mode: latex
%%% TeX-master: "root"
%%% End:

% LocalWords:  Acknowledgements Alwen Tiu Baelde LICS REUSSI CCR Equipes Associ

\bibliographystyle{plainnat}
%\bibliography{../references/master}

\begin{thebibliography}{43}
\providecommand{\natexlab}[1]{#1}
\providecommand{\url}[1]{\texttt{#1}}
\expandafter\ifx\csname urlstyle\endcsname\relax
  \providecommand{\doi}[1]{doi: #1}\else
  \providecommand{\doi}{doi: \begingroup \urlstyle{rm}\Url}\fi

\bibitem[Aydemir et~al.(2008)Aydemir, Chargu\'eraud, Pierce, Pollack, and
  Weirich]{aydemir08popl}
Brian Aydemir, Arthur Chargu\'eraud, Benjamin~C. Pierce, Randy Pollack, and
  Stephanie Weirich.
\newblock Engineering formal metatheory.
\newblock In \emph{35th ACM Symp.\ on Principles of Programming Languages},
  pages 3--15. ACM, January 2008.

\bibitem[Aydemir et~al.(2005)Aydemir, Bohannon, Fairbairn, Foster, Pierce,
  Sewell, Vytiniotis, Washburn, Weirich, and Zdancewic]{aydemir05tphols}
Brian~E. Aydemir, Aaron Bohannon, Matthew Fairbairn, J.~Nathan Foster,
  Benjamin~C. Pierce, Peter Sewell, Dimitrios Vytiniotis, Geoffrey Washburn,
  Stephanie Weirich, and Steve Zdancewic.
\newblock Mechanized metatheory for the masses: The {POPLmark} challenge.
\newblock In \emph{Theorem Proving in Higher Order Logics: 18th International
  Conference}, number 3603 in LNCS, pages 50--65. Springer-Verlag, 2005.

\bibitem[Baelde(2008)]{baelde08phd}
David Baelde.
\newblock \emph{A linear approach to the proof-theory of least and greatest
  fixed points}.
\newblock PhD thesis, Ecole Polytechnique, December 2008.
\newblock URL \url{http://www.lix.polytechnique.fr/~dbaelde/thesis/}.

\bibitem[Baelde et~al.(2007)Baelde, Gacek, Miller, Nadathur, and
  Tiu]{baelde07cade}
David Baelde, Andrew Gacek, Dale Miller, Gopalan Nadathur, and Alwen Tiu.
\newblock The {Bedwyr} system for model checking over syntactic expressions.
\newblock In F.~Pfenning, editor, \emph{21th Conf.\ on Automated Deduction
  (CADE)}, number 4603 in LNAI, pages 391--397. Springer, 2007.
\newblock URL
  \url{http://www.lix.polytechnique.fr/Labo/Dale.Miller/papers/cade2007.pdf}.

\bibitem[Bertot and Cast\'eran(2004)]{bertot04book}
Yves Bertot and Pierre Cast\'eran.
\newblock \emph{Interactive Theorem Proving and Program Development. Coq'Art:
  The Calculus of Inductive Constructions}.
\newblock Texts in Theoretical Computer Science. Springer Verlag, 2004.
\newblock URL \url{http://www.labri.fr/publications/l3a/2004/BC04}.

\bibitem[Church(1940)]{church40}
Alonzo Church.
\newblock A formulation of the simple theory of types.
\newblock \emph{J. of Symbolic Logic}, 5:\penalty0 56--68, 1940.

\bibitem[Coquand and Paulin(1988)]{coquand88colog}
Thierry Coquand and Christine Paulin.
\newblock Inductively defined types.
\newblock In \emph{Conference on Computer Logic}, volume 417 of \emph{LNCS},
  pages 50--66. Springer-Verlag, 1988.

\bibitem[Despeyroux et~al.(1995)Despeyroux, Felty, and
  Hirschowitz]{despeyroux95tlca}
Jo{\"{e}}lle Despeyroux, Amy Felty, and Andre Hirschowitz.
\newblock Higher-order abstract syntax in {Coq}.
\newblock In \emph{Second International Conference on Typed Lambda Calculi and
  Applications}, pages 124--138, April 1995.

\bibitem[Felty and Momigliano(2009)]{felty09ppdp}
Amy Felty and Alberto Momigliano.
\newblock Reasoning with hypothetical judgments and open terms in {H}ybrid.
\newblock In \emph{ACM SIGPLAN Conference on Principles and Practice of
  Declarative Programming (PPDP)}, pages 83--92, 2009.

\bibitem[Felty and Momigliano(2010)]{felty10jar}
Amy Felty and Alberto Momigliano.
\newblock Hybrid: {A} definitional two-level approach to reasoning with
  higher-order abstract syntax.
\newblock \emph{Journal of Automated Reasoning}, 2010.
\newblock To appear.

\bibitem[Gacek(2008)]{gacek08ijcar}
Andrew Gacek.
\newblock The {A}bella interactive theorem prover (system description).
\newblock In A.~Armando, P.~Baumgartner, and G.~Dowek, editors, \emph{Fourth
  International Joint Conference on Automated Reasoning}, volume 5195 of
  \emph{LNCS}, pages 154--161. Springer, 2008.
\newblock URL \url{http://arxiv.org/abs/0803.2305}.

\bibitem[Gacek(2009{\natexlab{a}})]{abella.website}
Andrew Gacek.
\newblock The {A}bella system and homepage.
\newblock \url{http://abella.cs.umn.edu/}, 2009{\natexlab{a}}.

\bibitem[Gacek(2009{\natexlab{b}})]{gacek09phd}
Andrew Gacek.
\newblock \emph{A Framework for Specifying, Prototyping, and Reasoning about
  Computational Systems}.
\newblock PhD thesis, University of Minnesota, 2009{\natexlab{b}}.

\bibitem[Gacek et~al.(2008)Gacek, Holte, Nadathur, Qi, and
  Snow]{teyjus.website}
Andrew Gacek, Steven Holte, Gopalan Nadathur, Xiaochu Qi, and Zach Snow.
\newblock The {T}eyjus system -- version 2, March 2008.
\newblock Available from \url{http://teyjus.cs.umn.edu/}.

\bibitem[Gacek et~al.(2009)Gacek, Miller, and Nadathur]{gacek09corr}
Andrew Gacek, Dale Miller, and Gopalan Nadathur.
\newblock Nominal abstraction.
\newblock Technical report, CoRR, August 2009.
\newblock URL \url{http://arxiv.org/abs/0908.1390}.
\newblock Extended version LICS 2008 paper. Submitted.

\bibitem[Harper et~al.(1993)Harper, Honsell, and Plotkin]{harper93jacm}
Robert Harper, Furio Honsell, and Gordon Plotkin.
\newblock A framework for defining logics.
\newblock \emph{Journal of the ACM}, 40\penalty0 (1):\penalty0 143--184, 1993.

\bibitem[Kahn(1987)]{kahn87stacs}
Gilles Kahn.
\newblock Natural semantics.
\newblock In \emph{Proceedings of the Symposium on Theoretical Aspects of
  Computer Science}, volume 247 of \emph{LNCS}, pages 22--39. Springer, March
  1987.

\bibitem[Landin(1964)]{landin64}
P.~J. Landin.
\newblock The mechanical evaluation of expressions.
\newblock \emph{Computer Journal}, 6\penalty0 (5):\penalty0 308--320, 1964.

\bibitem[Licata et~al.(2008)Licata, Zeilberger, and Harper]{licata08lics}
Daniel~R. Licata, Noam Zeilberger, and Robert Harper.
\newblock Focusing on binding and computation.
\newblock In F.~Pfenning, editor, \emph{23th Symp.\ on Logic in Computer
  Science}, pages 241--252. IEEE Computer Society Press, 2008.

\bibitem[McDowell and Miller(2000)]{mcdowell00tcs}
Raymond McDowell and Dale Miller.
\newblock Cut-elimination for a logic with definitions and induction.
\newblock \emph{Theoretical Computer Science}, 232:\penalty0 91--119, 2000.

\bibitem[McDowell and Miller(2002)]{mcdowell02tocl}
Raymond McDowell and Dale Miller.
\newblock Reasoning with higher-order abstract syntax in a logical framework.
\newblock \emph{ACM Trans.\ on Computational Logic}, 3\penalty0 (1):\penalty0
  80--136, 2002.

\bibitem[Miller(1992)]{miller92jsc}
Dale Miller.
\newblock Unification under a mixed prefix.
\newblock \emph{Journal of Symbolic Computation}, 14\penalty0 (4):\penalty0
  321--358, 1992.

\bibitem[Miller(2000)]{miller00cl}
Dale Miller.
\newblock Abstract syntax for variable binders: An overview.
\newblock In John Lloyd and {\em et al.}, editors, \emph{Computational Logic -
  {CL} 2000}, number 1861 in LNAI, pages 239--253. Springer, 2000.
\newblock URL
  \url{http://www.lix.polytechnique.fr/Labo/Dale.Miller/papers/ltrees.pdf}.

\bibitem[Miller and Tiu(2005)]{miller05tocl}
Dale Miller and Alwen Tiu.
\newblock A proof theory for generic judgments.
\newblock \emph{ACM Trans.\ on Computational Logic}, 6\penalty0 (4):\penalty0
  749--783, October 2005.
\newblock URL
  \url{http://www.lix.polytechnique.fr/Labo/Dale.Miller/papers/tocl-nabla.pdf}.

\bibitem[Miller et~al.(1991)Miller, Nadathur, Pfenning, and
  Scedrov]{miller91apal}
Dale Miller, Gopalan Nadathur, Frank Pfenning, and Andre Scedrov.
\newblock Uniform proofs as a foundation for logic programming.
\newblock \emph{Annals of Pure and Applied Logic}, 51:\penalty0 125--157, 1991.

\bibitem[Milner(1992)]{milner92mscs}
Robin Milner.
\newblock Functions as processes.
\newblock \emph{Mathematical Structures in Computer Science}, 2:\penalty0
  119--141, 1992.

\bibitem[Nadathur and Miller(1988)]{nadathur88iclp}
Gopalan Nadathur and Dale Miller.
\newblock An {Overview} of {$\lambda$Prolog}.
\newblock In \emph{{Fifth International Logic Programming Conference}}, pages
  810--827, Seattle, August 1988. MIT Press.
\newblock URL
  \url{http://www.lix.polytechnique.fr/Labo/Dale.Miller/papers/iclp88.pdf}.

\bibitem[Nadathur and Mitchell(1999)]{nadathur99cade}
Gopalan Nadathur and Dustin~J. Mitchell.
\newblock System description: {Teyjus} --- {A} compiler and abstract machine
  based implementation of {$\lambda$Prolog}.
\newblock In H.~Ganzinger, editor, \emph{16th Conf.\ on Automated Deduction
  (CADE)}, number 1632 in LNAI, pages 287--291, Trento, 1999. Springer.

\bibitem[Nipkow et~al.(2002)Nipkow, Paulson, and Wenzel]{nipkow02book}
Tobias Nipkow, Lawrence~C. Paulson, and Markus Wenzel.
\newblock \emph{Isabelle/{HOL}: {A} Proof Assistant for Higher-Order Logic}.
\newblock Springer, 2002.
\newblock LNCS Tutorial 2283.

\bibitem[Pfenning and Sch{\"u}rmann(1999)]{pfenning99cade}
Frank Pfenning and Carsten Sch{\"u}rmann.
\newblock System description: Twelf --- {A} meta-logical framework for
  deductive systems.
\newblock In H.~Ganzinger, editor, \emph{16th Conf.\ on Automated Deduction
  (CADE)}, number 1632 in LNAI, pages 202--206, Trento, 1999. Springer.

\bibitem[Pientka(2008)]{pientka08popl}
Brigitte Pientka.
\newblock A type-theoretic foundation for programming with higher-order
  abstract syntax and first-class substitutions.
\newblock In \emph{35th Annual {ACM} Symposium on Principles of Programming
  Languages (POPL'08)}, pages 371--382. ACM, 2008.

\bibitem[Pitts(2003)]{pitts03ic}
Andrew~M. Pitts.
\newblock Nominal logic, {A} first order theory of names and binding.
\newblock \emph{Information and Computation}, 186\penalty0 (2):\penalty0
  165--193, 2003.

\bibitem[Plotkin(1976)]{plotkin76}
Gordin Plotkin.
\newblock Call-by-name, call-by-value and the $\lambda$-calculus.
\newblock \emph{Theoretical Computer Science}, 1\penalty0 (1):\penalty0
  125--159, 1976.

\bibitem[Plotkin(1977)]{plotkin77}
Gordon Plotkin.
\newblock {LCF} as a programming language.
\newblock \emph{Theoretical Computer Science}, 5, 1977.

\bibitem[Plotkin(1981)]{plotkin81}
Gordon Plotkin.
\newblock A structural approach to operational semantics.
\newblock {DAIMI} {FN}-19, Aarhus University, Aarhus, Denmark, September 1981.

\bibitem[Poswolsky and Sch{\"u}rmann(2008)]{poswolsky08lfmtp}
Adam Poswolsky and Carsten Sch{\"u}rmann.
\newblock System description: Delphin - {A} functional programming language for
  deductive systems.
\newblock In A.~Abel and C.~Urban, editors, \emph{International Workshop on
  Logical Frameworks and Meta-Languages: Theory and Practice (LFMTP 2008)},
  volume 228, pages 113--120, 2008.

\bibitem[Reynolds(1972)]{reynolds72acm}
John Reynolds.
\newblock Definitional interpreters for higher order programming languages.
\newblock In \emph{ACM Conference Proceedings}, pages 717--740. ACM, 1972.

\bibitem[Sangiorgi(1994)]{sangiorgi94ic}
Davide Sangiorgi.
\newblock The lazy lambda calculus in a concurrency scenario.
\newblock \emph{Information and Computation}, 111\penalty0 (1):\penalty0
  120--153, May 1994.

\bibitem[Sch{\"{u}}rmann(2000)]{schurmann00phd}
Carsten Sch{\"{u}}rmann.
\newblock \emph{Automating the Meta Theory of Deductive Systems}.
\newblock PhD thesis, Carnegie Mellon University, October 2000.
\newblock URL \url{http://www.cs.yale.edu/homes/carsten/papers/S00b.ps.gz}.
\newblock CMU-CS-00-146.

\bibitem[Smorynski(2004)]{smorynski04hpl}
Craig Smorynski.
\newblock Modal logic and self-reference.
\newblock In Dov Gabbay and Franz Guenther, editors, \emph{Handbook of
  Philosophical Logic, Volume 11 (Second Edition)}, pages 1--54. Kluwer
  Academic, 2004.

\bibitem[Tiu(2004)]{tiu04phd}
Alwen Tiu.
\newblock \emph{A Logical Framework for Reasoning about Logical
  Specifications}.
\newblock PhD thesis, Pennsylvania State University, May 2004.
\newblock URL
  \url{http://etda.libraries.psu.edu/theses/approved/WorldWideIndex/ETD-479/}.

\bibitem[Tiu(2006)]{tiu06lfmtp}
Alwen Tiu.
\newblock A logic for reasoning about generic judgments.
\newblock In A.~Momigliano and B.~Pientka, editors, \emph{Int. Workshop on
  Logical Frameworks and Meta-Languages: Theory and Practice (LFMTP'06)}, 2006.

\bibitem[Urban(2008)]{urban08jar}
Christian Urban.
\newblock Nominal reasoning techniques in {I}sabelle/{HOL}.
\newblock \emph{Journal of Automated Reasoning}, 40\penalty0 (4):\penalty0
  327--356, 2008.

\end{thebibliography}
%% Copy of root.bbl with URLs removed except those from arvix or
%% for systems available online

%%% Local Variables: 
%%% mode: latex
%%% TeX-master: "root"
%%% End: 

\end{document}